\documentclass[journal,twoside]{IEEEtran}
\usepackage{mathrsfs}
\usepackage{amsbsy}
\usepackage{graphicx}
\usepackage{xcolor}
\usepackage{amsmath}
\usepackage{amsthm}
\usepackage{amssymb}

\usepackage{nomencl}
\usepackage{commath}
\usepackage[T1]{fontenc}
\usepackage{mathtools}
\usepackage{float}
\usepackage{url}
\usepackage{algorithm}
\usepackage{pifont}
\usepackage{algpseudocode}
\usepackage{etoolbox}
\usepackage{lipsum}

\usepackage{subfigure}
\usepackage{hhline}
\usepackage{multirow}

\usepackage[shortlabels]{enumitem}
\usepackage{blindtext}
\usepackage{wasysym}
\usepackage{balance}
\usepackage{array}
\usepackage{bm}  
\usepackage{makecell}

\makenomenclature

\usepackage{cite}
\ifCLASSINFOpdf
\else
\fi
\begin{document}
\title{Robust Economic Dispatch with Flexible Demand and Adjustable Uncertainty Set}

%~\IEEEmembership{Student Member,~IEEE,} ~\IEEEmembership{Member,~IEEE,} ~\IEEEmembership{Fellow,~IEEE}

\author{Tian~Liu,~\IEEEmembership{Student Member,~IEEE,}\thanks{T. Liu, S. Wang, and D.H.K. Tsang are with the Hong Kong University of Science and Technology, Hong Kong (e-mail: \{tliuai, swangbr, eetsang\}@ust.hk). 
} Su~Wang,~\IEEEmembership{Student Member,~IEEE,}  and~Danny~H.K.~Tsang,~\IEEEmembership{Fellow,~IEEE}}

%\thanks{M. Shell is with the Department
%of Electrical and Computer Engineering, Georgia Institute of Technology, Atlanta,
%\thanks{Manuscript received April 19, 2005; revised September 17, 2014.}}


%Journal of \LaTeX\ Class Files,~Vol.~13, No.~9, September~2014

\markboth{}{}




\maketitle

\begin{abstract}
With more renewable energy sources (RES) integrated into the power system, the intermittency of RES places a heavy burden on the system. The uncertainty of RES is traditionally handled by controllable generators to balance the real time wind power deviation. As the demand side management develops, the flexibility of aggregate loads can be leveraged to mitigate the negative impact of the wind power. In view of this, we study the problem of how to exploit the multi-dimensional flexibility of elastic loads to balance the trade-off between a low generation cost and a low system risk related to the wind curtailment and the power deficiency. These risks are captured by the conditional value-at-risk. Also, unlike most of the existing studies, the uncertainty set of the wind power output in our model is not fixed. By contrast, it is undetermined and co-optimized based on the available load flexibility. We transform the original optimization problem into a convex one using surrogate affine approximation such that it can be solved efficiently. In case studies, we apply our model on a six-bus transmission network and demonstrate that how flexible load aggregators can help to determine the optimal admissible region for the wind power deviations.

\end{abstract}


% \fi
\IEEEpeerreviewmaketitle

\section{Introduction}

%		 \begin{enumerate}[1)]
%		 	\item Our framework is generic and not limited to OPF problems because we directly approximate the rank function, which is not problem specific at all. In fact, any problem that can be equivalently formulated into QCQPs can be solved by the proposed 
%		 \end{enumerate} 
%	 With numerical results, we have applied our method to the OPF related problems and show that it can find the real optimal point when the duality gap is zero and even for non-zero case, it can find the solution as good as some commercial solver.
\IEEEPARstart{A}{s} the integration of renewable energy sources (RESs) into the power grid, such as solar and wind power, is ever-increasing, these is a growing demand for the flexible resources to mitigate the negative impacts induced by the intermittent nature of RESs. The flexible resources can respond to the changes in net load, where net load is defined as the system load not served by RESs \cite{lannoye2012evaluation}. Potential flexible resources can be generators that can ramp quickly, demand responsive loads \cite{cecati2011combined} and energy storage, etc.

With the help of flexible resources, the power system can resist the variance of the RES outputs to a certain extent. For the uncertainty that can not be accommodated by the system, loss will be incurred \cite{albadi2010overview}. For example, the RES output has to be curtailed in case of over-generation and the load has to be discarded in case of under-generation. In spite of this, it is hardly possible that the power system can be guaranteed to be immune to any possible realization of the RES outputs, since it will require a tremendous amount of flexible recourses, which will be, however, very costly. Therefore, the system needs to consider the trade-off between the system cost and the risk of loss. In the traditional day-ahead energy scheduling, the system operator aims at minimizing the generation cost while satisfying the operational constraints such as the load balance constraints and the transmission line capacity constraints, etc. When the RES is taken into account, the flexibility of the system, namely the potential capability of the system to cope with the variability and uncertainty \cite{ackermann2005wind,li2018flexible}, also needs to be determined. 

There are several works in the literature that have studied the optimization of the wind power uncertainty set in the economic dispatch problem or the unit commitment problem \cite{li2018flexible,wei2016dispatchability,wang2017robust,ye2018surrogate,wang2016risk}. In \cite{li2018flexible}, the authors study the look-ahead dispatch (LAD) and propose to manage operational uncertainties over the next several hours by utilizing the load and intermittent generation forecasts while incorporating the inter-temporal constraint considering large-scale wind power integration. A flexible LAD model is developed to balance the operational costs and the conditional value-at-risk of wind power (CVaR-WP) based on robust optimization (RO). According to the proposed model, the base points, participation factors, and flexible capacity of automatic generation control units are co-optimized. In addition, a reasonable admissible region of wind power (ARWP) on each node can be obtained correspondingly. An approach based on the big-M method and a decomposition method is presented to improve the computing efficiency. In \cite{wei2016dispatchability}, Wei \textit{et al.} consider the dispatchability as the set of all admissible nodal wind power injections that will not cause
infeasibility in real-time dispatch (RTD) and propose two mathematical formulations of the dispatchability maximized energy and reserve dispatch (DM-ERD). Efficient convex optimization based algorithms are developed to solve these two models. Different from the conventional robust optimization method, their model does not rely on the specific uncertainty set of wind generation and directly optimizes the uncertainty accommodation capability of the power system. In \cite{wang2017robust}, a robust risk-constrained unit commitment (RRUC) formulation is proposed to cope with large-scale volatile and uncertain wind generation, which addresses the issue that how large the prescribed uncertainty set should be when it is used for robust unit commitment (RUC) decision making. Compared with RUC, the boundaries of wind generation uncertainty set are adjustable variables to be optimized in RRUC, resulting in an optimal allocation of operational flexibility as well as operational risk mitigating capability. Ye \textit{et al.} \cite{ye2018surrogate} present an approach to co-optimize transactive flexibility, energy, and optimal injection-range of
Variable Energy Resource (VER). Their approach proactively positions the flexible resources and optimizes the demand of flexibility, in which flexibility is defined as the change range of power injection that the system can accommodate using available flexible resources within a specified time. A novel surrogate affine approximation method is proposed to solve the problem in polynomial time. It is proved that its solution is even better than the original affine policy-based method used in the power literature.

It is noted that in the literature above, in order to provide the flexibility to the system, only fast reacting generators are leveraged, which is not cost efficient and those generators have limited flexibility capacity. By contrast, the flexibility of the demand side is a promising resource to be exploited. There are works in the literature trying to handle the uncertainty of RESs using demand side management \cite{bukhsh2016integrated,wu2015thermal,wang2015optimal,zhao2012robust,falsafi2014role}. In \cite{bukhsh2016integrated}, Bukhsh \textit{et al.} present a two-stage stochastic programming approach to solve a multiperiod optimal power flow problem under renewable generation uncertainty. In that work, the operating points of the conventional power plants are determined in the first stage and the second stage optimally accommodates the realized generation from the renewable resources relying on the demand-side flexibility. They observe that considerable savings in power generation costs can be made if a small proportion of the demand is flexible. In \cite{wu2015thermal}, Wu \textit{et al.} propose a stochastic day-ahead scheduling of electric power systems with flexible resources, which include thermal units with up/down ramping capability, energy storage, and hourly demand response (DR), for managing the variability of RESs. They show that the hourly DR minimizes the expected level of variability, yielding a more flat net load profile for thermal units. In \cite{wang2015optimal}, Wang \textit{et al.} use demand response to manage wind power intermittency by shifting the time that electrical power system loads occur in response to real-time prices and wind availability. They propose an optimization model for the economic dispatch of a transmission constrained system with a high penetration of wind power and determine the optimal sizing and distribution of DR given a fixed budget for customer
incentives and the installation of enabling technology.

Although the demand side flexibility is considered in the aforementioned works, there exist several issues. One is that the system is required to accommodate a predefined uncertainty set, while the optimization of the admissible region for wind powers is not incorporated in their works. Another issue is that those works typically consider a certain aspect of the demand flexibility, such as the shifting flexibility for deferrable loads. In fact, the load flexibility can have multiple dimensions referring to various aspects of the flexibility \cite{liu2017market}. For example, i) the shifting flexibility \cite{logenthiran2012demand}, which means that a certain load can be shifted to a later or an earlier time point; ii) the energy flexibility, which refers to that the electricity consumers, such as electric vehicles (EVs), can adjust their energy demands for economic benefits \cite{sun2018eliciting}; iii) power flexibility \cite{negrete2018rate}, which considers the fact that the power rating of the energy consumer at each time slot is restricted within a certain range. Other dimension of the flexibility can be the deadline flexibility. An example of this is the EVs that are not sensitive to the charging deadlines and hence are capable of delaying their charging request up to some extended deadlines \cite{sun2018eliciting}. 

Complementary to existing works, this paper proposes a unified framework to integrate the load aggregators with multi-dimensional flexibility (MDF) into the day-ahead economic dispatch problem with uncertain wind powers. The load flexibility is exploited to accommodate the uncertainty of the wind power outputs. However, because of the limited flexibility that can be offered, the system can only resist a certain level of variability of wind at each node in the network. For those realizations of wind power outputs beyond the uncertainty set that the system can deal with, additional system loss is incurred, which is modeled by the CVaR for both the wind curtailment loss as well as the power deficiency. The benefit of using the CVaR to quantify the system risk is that comparing to traditional robust optimization technique, the probabilistic characteristic of the wind power can be captured \cite{rockafellar2002conditional,rockafellar2000optimization}. Moreover, the uncertainty set of the wind power outputs that the system can accommodate is not predetermined, but regarded as decision variables to be co-optimized. Therefore, we are trying to balance the trade-off between a low operational cost and a low risk for the system loss by simultaneously optimizing the base points for the generator outputs, the scheduling of the flexible loads as well as the admissible uncertainty set for the wind power outputs on each bus at different time slots. 

In order to solve the proposed optimization problem, we first transform the CVaR related objective function to a convex one using the technique in \cite{rockafellar2000optimization}. Then, since the uncertainty set incorporates decision variables, it will render the problem non-convex if the traditional affine policy is used for allocating the deviation of the wind power outputs to the load aggregators. In this regard, the surrogate affine approximation (SAA) method \cite{ye2018surrogate} is adopted, which can transform the original problem into a convex one such that it can be solved efficiently. In summary, the contributions of this paper are twofold as follows:

\begin{enumerate}
	\item We propose a unified framework to integrate the load aggregators with MDF into the economic dispatch problem under wind uncertainty. The MDF is leveraged to offer the flexibility to the system such that it can accommodate a certain admissible uncertainty set for the wind power outputs. The wind power uncertainty set is also part of the decision variables. Thus, our model can co-optimize the admissible uncertainty set of wind power using the load flexibility while taking the generation cost and the CVaR-related system loss into account.
	\item We reformulate our original optimization problem into a convex one by: i) transforming the CVaR calculation into solving a linear program and ii) applying the SAA method to handle the robust constraints with decision variables in the uncertainty set, which is non-convex if the traditional affine policy is adopted.

\end{enumerate}

The rest of this paper is organized as follows. In Section \ref{j4:sec2}, the system modeling is presented. Then, the problem formulation is described in Section \ref{j4:sec3}. The case studies are given in Section \ref{j4:sec4},  followed by the conclusion in Section \ref{j4:sec5}.

\section{System Modeling}\label{j4:sec2}
\subsection{Controllable Generator and Fixed Load Model}
We consider a transmission network with $N$ buses and $M$ transmission lines. The time horizon is discretized, which is denoted by ${\cal T}=\{1,2,\cdots, T\}$, and $T$ is the total number of time slots. Denote the set of all buses as $\mathcal{N}=\{1,2,\cdots,N\}$ and the set of all transmission lines as ${\cal M}=\{1,2,\cdots,M\}$. Suppose there are $n_g$ controllable generators located at buses in ${\cal N}^G \subseteq {\cal N}$, and $\boldsymbol{g}_t=\{g_{i,t}, i \in \cal N^G\}$ denotes the vector of power outputs from these generators at time $t$. They should satisfy the following constraints:
\begin{equation}
	\boldsymbol{g}^- \leq \boldsymbol{g}_t \leq \boldsymbol{g}^+, \ \forall t \in \cal T,
\end{equation}
where $\boldsymbol{g}^-$ and $\boldsymbol{g}^+$ denote the vectors of lower and upper bounds for controllable generators, respectively. In addition, the set of buses with fixed loads is denoted by $\cal N^D \subseteq \cal N$ and $\boldsymbol{d}_t=\{d_{i,t}, i \in \cal N^D\}$ represents the vector of fixed loads at time $t$.

\subsection{Load Aggregators with MDF}
We denote the bus set of load aggregators with MDF by $\cal N^F \subseteq \cal N$, and the cardinality of $\cal{N}^F$ is $n_f$. Also, the vector of flexible load power consumptions at time $t$ is denoted by $\boldsymbol{x}_t=\{x_{i,t}, i \in \cal N^F\}$. For each load aggregator, we denote the column vector of power consumption across the total time horizon as $\boldsymbol{x}_{\cdot i}=\{x_{i,t}, t \in {\cal T}\}$. Then, the flexible region of the power consumption for each load aggregator $i$ is described by the following constraints:
\begin{equation}
	\bm{l}_{ i}\leq \bm{L} {\bm{x}_{\cdot i}} \leq\bm{u}_{ i}, \ \forall i \in \cal N^F, \label{wp-mdf:2}
\end{equation} 
\begin{equation}
	\bm{x}_{ i}^- \leq {\bm{x}_{\cdot i}} \leq \bm{x}_{ i}^+,  \ \forall i \in \cal N^F,\label{wp-mdf:3}
\end{equation}
where $\bm{l}_i$ and $\bm{u}_i$ denote the lower and upper bounds for the cumulative energy consumption from $t = 1 $ up to each time slot $t, t\in \cal T$, respectively. Moreover, $\bm{x}^-_i$ and $\bm{x}^+_i$ denote the lower and upper bounds for the power consumption at each time slot, respectively. $\bm{L}$ is a lower triangular matrix with all the non-zero elements equal to one. Note that constraints \eqref{wp-mdf:2} and \eqref{wp-mdf:3} are general enough to represent the flexible loads with MDF. Specifically, the shifting flexibility and energy flexibility can be captured by constraint (\ref{wp-mdf:2}) while the power flexibility can be represented by constraint \eqref{wp-mdf:3}.

\subsection{Wind Power With Uncertainty}
	Denote the set of buses with wind generators as $\cal N^W \subseteq \cal N$ and the cardinality of ${\cal N}^W$ is $n_w$. Then, the vector of wind power outputs at time $t$, $\bm{w}_t=\{w_{i,t}, i \in \cal N^W\}$, is represented by
\begin{equation}
	\bm{w}_t = \overline{\bm{w}}_t+\bm{\delta}_t, \ \bm{\delta}_t \in [-\bm{\delta}_t^-, \bm{\delta}_t^+],
\end{equation}
where $\overline{\bm{w}}_t$ is the vector of the wind power forecast values and $\bm{\delta}_t =\{\delta_{i,t}, i \in \cal N^W\}$ denotes the uncertain deviation of the actual power from the forecast ones. Suppose the wind power deviation is bounded by the lower and upper bounds, $-\bm{\delta}_t^- $ and $\bm{\delta}_t^+$, respectively. Note that both  $\bm{\delta}_t^- (\geq 0)$ and $\bm{\delta}_t^+ (\geq 0)$ are decision variables, such that the system can optimally determine the uncertainty level which can be accommodated considering the system cost and risk simultaneously. When the wind power outputs are not as forecasted, i.e., $\bm{\delta}_t$ is non-zero, then, the system will rely on the load aggregators with MDF to handle the uncertainty. Specifically, the actual power consumption vector of flexible loads is defined as
\begin{equation}
	\bm{x}_t = \overline{\bm{x}}_t +\Delta \bm{x}_t(\bm{\delta}_t),
\end{equation} 
where $ \overline{\bm{x}}_t$ denotes the set-points for the flexible loads corresponding to forecast wind powers and $\Delta \bm{x}_t(\bm{\delta}_t)$ denotes the vector of power adjustments with respect to the wind power deviations. Thus, we leverage the load flexibility to accommodate the wind power uncertainty.

\subsection{Conditional Value-at-Risk}
In our problem, for all the wind power deviations within the uncertainty set $\bm{\delta}_t \in [-\bm{\delta}_t^-, \bm{\delta}_t^+]$, the system will resort to the flexible loads to deal with it. However, when the deviation is beyond the set that the system is able to handle, loss will be incurred. We use CVaR to capture such loss. Specifically, by defining the operator $[x]^+=\text{max}(x,0),$ we define two loss functions as
\begin{align}
	& f^1(\delta,w)_{i,t}=[w_{i,t}-(\overline{w}_{i,t}+\delta_{i,t}^+)]^+,\label{j4:cvar-loss-1} \\
	& f^2(\delta,w)_{i,t}=[(\overline{w}_{i,t}-\delta_{i,t}^-)-w_{i,t}]^+, 
\end{align}
where $f^1(\delta,w)_{i,t}$ represents the wind power that will be curtailed when it is greater than the upper bound of the admissible uncertainty set, while $f^2(\delta,w)_{i,t}$ refers to the power deficiency when it is less than the lower bound. Suppose the probability distribution of $w_{i,t}$ is denoted by $p(w_{i,t})$, then the probability of $f^1(\delta,w)_{i,t}$ not exceeding a threshold $\alpha^1_{i,t}$ is
\begin{equation}
	\psi^1_{i,t}(\alpha^1_{i,t})=\int\displaylimits_{f^1(\delta,w)_{i,t} \leq \alpha^1_{i,t}}p(w_{i,t})d(w_{i,t}).
\end{equation}
According to the definition in \cite{rockafellar2000optimization}, the $\beta$-VaR$_{i,t}$ and $\beta$-CVaR$_{i,t}$ values are given by
\begin{equation}
	\alpha^{1,\beta}_{i,t}={\text {min}}\{\alpha^1_{i,t} \in \mathbb{R}: \psi_{i,t}^1(\alpha^1_{i,t}) \geq \beta \},
\end{equation}
and 
\begin{align}
	& \phi^{1,\beta}_{i,t}= (1-\beta)^{-1}\int\displaylimits_{f^1(\delta,w)_{i,t} \geq \alpha^{1,\beta}_{i,t}}f^1(\delta,w)_{i,t}p(w_{i,t})d(w_{i,t}). \label{j4:cvar-1}
\end{align}
The meaning of $\beta{\text -}\textrm{CVaR}_{i,t}$ is the conditional expectation of the loss given that the loss is greater than or equal to $\beta{\text -}\textrm{VaR}_{i,t}$.
Similarly, the $\beta{\text -}\textrm{CVaR}_{i,t}$ value for the loss function $f^2(\delta,w)_{i,t}$ is given by
\begin{align}
		& \phi^{2,\beta}_{i,t}=(1-\beta)^{-1}\int\displaylimits_{f^2(\delta,w)_{i,t} \geq \alpha^{2,\beta}_{i,t}}f^2(\delta,w)_{i,t}p(w_{i,t})d(w_{i,t}).\label{j4:cvar-2}
\end{align}

\section{Problem Formulation}\label{j4:sec3}
\subsection{Multi-Period Economic Dispatch under Wind Uncertainty}
We denote the matrix of generation shift factors as ${\Gamma}\in \mathbb{R}^{M\times N}$. Also, we denote the bus connection matrix for buses with controllable generators, fixed loads, wind turbines and flexible load aggregators as $H_g$, $H_d$, $H_w$ and $H_f$, respectively. For example, $H_g \in \mathbb{R}^{N\times n_g}$ is constructed such that its $(i,j)$-th element is one if and only if generator $j$ is located at bus $i$, with all the other elements equal to zero. Moreover, we define the cost function for the controllable generators as
\begin{equation}
	C^G(g_{i,t})=c^2_{i}g^2_{i,t}+c^1_{i}g_{i,t}+c^0_{i},
\end{equation}
where $c^2_{i,t},c^1_{i,t}$ and $c^0_{i,t}$ are the quadratic, linear and constant coefficients, respectively.
Then the multi-period economic dispatch with undetermined wind uncertainty set is formulated as follows:
\begin{subequations}
\begin{align}
& {\text{min}} 
&& \sum_{t}\bigg(\sum_{i \in \cal{N}^G}\mathcal{C}^G({{g}_{i,t}})+ \sum_{i \in \cal{N}^W}\big(\eta^1 \phi^{1,\beta}_{i,t}+\eta^2 \phi^{2,\beta}_{i,t}\big)\bigg) \label{arwp:f1}\\
& \text{s.t.} &&\bm{1}^T(\bm{g}_{t}+{\overline{\bm{w}}_t}-\bm{d}_{t}-\overline{\bm{x}}_{t})=0,  \ \forall t ,\label{arwp:f2} \\
&&& |{\Gamma}(H_g{\bm{g}_{t}}+H_w\overline{\bm{w}_t}-H_d\bm{d}_{t}-H_f\overline{\bm{x}}_{t})|\leq\bm{q},  \ \forall t , \label{arwp:f3}\\
&&&\bm{g}_{t}^- \leq {\bm{g}_{t}} \leq \bm{g}_{t}^+, \ \forall t , \label{arwp:f4}\\
&&&\bm{l}_{ i}\leq \bm{L} \overline{\bm{x}}_{\cdot i} \leq\bm{u}_{ i}, \ \forall i \in\mathcal{N}^F, \label{arwp:f5} \\
&&&\bm{x}_{ i}^- \leq \overline{\bm{x}}_{\cdot i} \leq \bm{x}_{ i}^+,  \ \forall i \in\mathcal{N}^F, \label{arwp:f6} \\
%&&& \widetilde{\bm{w}_t}=\overline{\bm{w}_t}+{\color{red}\bm{\delta}_t}, \bm{\delta}_t \in [-\bm{\delta}_t^-,\bm{\delta}_t^+], \ \forall t , \\
&&& \bm{w}_t = \overline{\bm{w}}_t+\bm{\delta}_t, \ \forall t, \label{arwp:f7} \\
&&& \bm{x}_t = \overline{\bm{x}}_t +\Delta \bm{x}_t(\bm{\delta}_t), \ \forall t, \label{arwp:f8} \\
&&& \bm{\delta}^-_t \geq \bm{0}, \ \forall t ,  \label{arwp:f9} \\
&&& \bm{\delta}^+_t \geq \bm{0}, \ \forall t ,  \label{arwp:f10} \\
%&&& \widetilde{\bm{x}_t}=\overline{\bm{x}_t}+{\color{red}\Delta\bm{x}_t(\bm{\epsilon_t})},  \  \forall t , \\
&&&\bm{l}_{ i}\leq \bm{L} {\bm{x}_{\cdot i}} \leq\bm{u}_{ i},  \ \forall i \in\mathcal{N}^F,  \forall \bm{\delta}_t \in [-\bm{\delta}_t^-, \bm{\delta}_t^+], \label{arwp:f11}\\
&&&\bm{x}_{ i}^- \leq {\bm{x}_{\cdot i}} \leq \bm{x}_{ i}^+,  \ \forall i \in\mathcal{N}^F,  \forall \bm{\delta}_t \in [-\bm{\delta}_t^-, \bm{\delta}_t^+], \label{arwp:f12} \\
&&& \bm{1}^T(\bm{g}_{t}{+}{{\bm{w}}_t}{-}\bm{d}_{t}{-}{\bm{x}_{t}}){=}0,  \forall t , \forall \bm{\delta}_t {\in} [-\bm{\delta}_t^-, \bm{\delta}_t^+],\label{arwp:f13} \\
&&& |{\Gamma}(H_g{\bm{g}_{t}}+H_w{\bm{w}_t}-H_d\bm{d}_{t}-H_f{\bm{x}_{t}})|\leq\bm{q}, \nonumber  \\
&&& \hspace{4.00cm} \forall t ,\forall \bm{\delta}_t {\in} [-\bm{\delta}_t^-, \bm{\delta}_t^+].\label{arwp:f14}
\end{align}
\end{subequations}
In this problem, the decision variables are $\bm{g}_t,\bm{x}_t,\overline{\bm{x}}_t,$ $\Delta \bm{x}_t(\bm{\delta}_t),\bm{\delta}^-_t$ and $\bm{\delta}^+_t, \forall t \in \cal T$. The objective is to minimize the weighted sum of the total generation cost plus the CVaR values corresponding to the wind curtailment risk and the power deficiency risk, where $\eta^1$ and $\eta^2$ are the weighting factors. Constraint \eqref{arwp:f2} requires that the fixed and flexible loads are balanced by the outputs from the wind turbines and the controllable generators at the set-points. Constraint \eqref{arwp:f3} means that the power flow on each transmission line does not violate the capacity in both directions, where $\bm{q}$ denotes the vector of transmission line capacities. Moreover, constraints \eqref{arwp:f5}--\eqref{arwp:f6} are the power consumption requirements for the flexible loads at the set-points. In addition, \eqref{arwp:f11}--\eqref{arwp:f14} are robust constraints for the flexible loads when the wind power deviation is within the uncertainty set to be determined. Note that this model is trying to help the operator find a balance between a low generation cost and a low risk level while satisfying all the operational constraints. Specifically, on the one hand, for large values of $\bm{\delta}^-_t$ and $\bm{\delta}^+_t$, the system will have low risks for the wind curtailment and the power deficiency. In spite of this, more flexibility will be reserved from the flexible load aggregators, which can increase the generation cost.

This problem is not easy to solve because of two difficulties. First, the CVaR values in the objective function is hard to minimize directly given its definition as \eqref{j4:cvar-1} and \eqref{j4:cvar-2}. Second, since the uncertainty set for constraints  \eqref{arwp:f11}--\eqref{arwp:f14} involves decision variables, this problem is not convex if the traditional affine policy{\footnote{The traditional affine policy means that the power adjustment for the flexible load is linearly related to the uncertain variables, i.e, for some constraint matrix $\bm{B},\Delta \bm{x}_t(\bm{\delta}_t)=\bm{B\delta_t}.$} is adopted. In the following, we will demonstrate how these two difficulties are handled such that the problem can be solved efficiently.

\subsection{Transformation of CVaR}
From \cite{rockafellar2000optimization}, it can be shown that the CVaR value $ \phi^{1,\beta}_{i,t}$ can be calculated by solving a minimization problem as follows:
\begin{equation}
		\phi^{1,\beta}_{i,t}= 	\underset{\gamma^1_{i,t} \in \mathbb{R}}{\textrm{min}} \ F^1_{\beta}(\gamma_{i,t}^1),
\end{equation}
where
\begin{equation}
	 F^1_{\beta}(\gamma_{i,t}^1)=\gamma^1_{i,t}+(1-\beta)^{-1}\int_{w_{i,t}}[f^1(\delta,w)_{i,t}-\gamma^1_{i,t}]^+p(w_{i,t})dw_{i,t}.\label{j4:cvar-trans-1}
\end{equation}
In \eqref{j4:cvar-trans-1}, $ F^1_{\beta}(\gamma_{i,t}^1)$ is convex and continuously differentiable according to Theorem 1 of \cite{rockafellar2000optimization}. Then the integral with respect to the random variable $w_{i,t}$ can be approximated using a data sample. Suppose that there are $K_{i,t}$ sample points, denoted as $w_{i,t}^k,i=1,2,\cdots,K_{i,t}$, sampled from the variable following the probability distribution of $w_{i,t}$. Then the approximation of \eqref{j4:cvar-trans-1} is given by
\begin{equation}
		\widetilde{F^1_{\beta}}(\gamma_{i,t}^1)= \gamma^1_{i,t}+\frac{1}{K_{i,t}(1-\beta)}\sum_{k=1}^{K_{i,t}}[f^1(\delta,w^k_{i,t})_{i,t}-\gamma^1_{i,t}]^+. \label{j4:cvar-trans-2}
\end{equation}
Therefore, the calculation of CVaR can be approximated by minimizing $\widetilde{F^1_{\beta}}(\gamma_{i,t}^1)$. In order to remove the $[\cdot]^+$ operator in \eqref{j4:cvar-trans-2}, we introduce the auxiliary real variables $v^{1,k}_{i,t}$ and the CVaR can be approximated by
\begin{equation}
		\underset{\gamma^1_{i,t} \in \mathbb{R}}{\textrm{min}} \ \widehat{F^1_{\beta}}(\gamma_{i,t}^1) \label{j4:cvar-obj-1}
\end{equation}
subject to
\begin{align}
		& v^{1,k}_{i,t} \geq 0, \label{j4:cvar-cstr-1}\\
		& v^{1,k}_{i,t}+\gamma^1_{i,t} \geq f^1(\delta,w^k_{i,t})_{i,t}, \forall \ k=1,\cdots, K_{i,t}, \label{j4:cvar-cstr-2}
\end{align}
where
\begin{equation}
	\widehat{F^1_{\beta}}(\gamma_{i,t}^1) = \gamma^1_{i,t}+\frac{1}{K_{i,t}(1-\beta)}\sum_{k=1}^{K_{i,t}}v^{1,k}_{i,t}. \label{j4:cvar-a1}
\end{equation}
Since there still exists the $[\cdot]^+$ operator in $f^1(\delta,w^k_{i,t})_{i,t}$ as defined in \eqref{j4:cvar-loss-1}, we introduce another auxiliary real variables $\mu^{1,k}_{i,t}$ such that constraints \eqref{j4:cvar-cstr-1}--\eqref{j4:cvar-cstr-2} are equivalent to the following constraints:
\begin{align}
&	v^{1,k}_{i,t}\geq 0,\label{j4:cvar-cstr-3}\\
& \mu^{1,k}_{i,t}\geq 0, \label{j4:cvar-cstr-4}\\
& v^{1,k}_{i,t}+\gamma^1_{i,t} \geq \mu^{1,k}_{i,t},\label{j4:cvar-cstr-5}\\
& \mu^{1,k}_{i,t} \geq w^k_{i,t}-(\overline{w}_{i,t}+\delta_{i,t}^+).\label{j4:cvar-cstr-6}
\end{align}
Therefore, CVaR can be obtained by solving the optimization problem with the objective of \eqref{j4:cvar-obj-1} subject to constraints \eqref{j4:cvar-cstr-3}--\eqref{j4:cvar-cstr-6}, which is a linear program. Similarly, $ \phi^{2,\beta}_{i,t}$ can be approximated by solving the following problem:
%\begin{equation}
%\label{j4:cvar-obj-2}
\begin{align}
& \underset{\gamma^2_{i,t} \in \mathbb{R}}{\textrm{min}} & & \widehat{F^2_{\beta}}(\gamma_{i,t}^2) \nonumber \\
& \text{s.t.} & &	v^{2,k}_{i,t}\geq 0,\label{j4:cvar-cstr-7}\\
&  & &\mu^{2,k}_{i,t}\geq 0, \label{j4:cvar-cstr-8}\\
& & & v^{2,k}_{i,t}+\gamma^2_{i,t} \geq \mu^{2,k}_{i,t},\label{j4:cvar-cstr-9}\\
& & & \mu^{2,k}_{i,t} \geq (\overline{w}_{i,t}-\delta_{i,t}^-)-w^k_{i,t},\label{j4:cvar-cstr-10}
\end{align}
%\end{equation}
where
\begin{equation}
\widehat{F^2_{\beta}}(\gamma_{i,t}^2) = \gamma^2_{i,t}+\frac{1}{K_{i,t}(1-\beta)}\sum_{k=1}^{K_{i,t}}v^{2,k}_{i,t}.\label{j4:cvar-a2}
\end{equation}
Therefore, we have shown that how to get CVaR values by solving linear programs and the original optimization problem \eqref{arwp:f1}--\eqref{arwp:f14} can be reformulated into
\begin{equation}
	{\text{min}} 
	\  \sum_{t}\bigg(\sum_{i \in \cal{N}^G}\mathcal{C}^G({{g}_{i,t}})+ \sum_{i \in \cal{N}^W}\big(\eta^1 \widehat{F^1_{\beta}}(\gamma_{i,t}^1)+\eta^2 \widehat{F^2_{\beta}}(\gamma_{i,t}^2)\big)\bigg), \label{arwp:fa1}
\end{equation}
subject to \eqref{arwp:f2}--\eqref{arwp:f14}, \eqref{j4:cvar-a1}--\eqref{j4:cvar-a2}, where the decision variables are $\bm{g}_t,\bm{x}_t,\overline{\bm{x}}_t,\Delta \bm{x}_t(\bm{\delta}_t),\bm{\delta}^-_t,\bm{\delta}^+_t,	v^{1,k}_{i,t},	v^{2,k}_{i,t}, \mu^{1,k}_{i,t}, \mu^{2,k}_{i,t},\gamma^1_{i,t}$ and $\gamma^2_{i,t}, \forall k=1,2,\cdots,K_{i,t},i \in {\cal{N}^W}, t \in \cal{ T}.$

\subsection{Surrogate Affine Approximation}
In this subsection, we introduce how to deal with constraints  \eqref{arwp:f11}--\eqref{arwp:f14} using the SAA method proposed in \cite{ye2018surrogate}. First, we group all the decision variables except $\Delta \bm{x}_t(\bm{\delta}_t),t\in \cal T$ as a column vector $\bm{y}$. Second, we denote the concatenation of $\Delta \bm{x}_t(\bm{\delta}_t)$ and $\bm{\delta}_t$ as $\Delta \bm{x}(\bm{\delta})=[\Delta \bm{x}_1(\bm{\delta}_1)',\cdots,\Delta \bm{x}_t(\bm{\delta}_T)']'$ and $\bm{\delta}=[\bm{\delta}_1',\cdots,\bm{\delta}_T']'$, respectively. Moreover, we use $\bm{\delta}^-=[(\bm{\delta}^{-}_1)',\cdots,(\bm{\delta}^{-}_T)']$ and $\bm{\delta}^+=[(\bm{\delta}^{+}_1)',\cdots,(\bm{\delta}^{+}_T)']'$ to group $\bm{\delta}^-_t$ and $\bm{\delta}_t^+$ across the entire time horizon, respectively. Then, constraints \eqref{arwp:f11}--\eqref{arwp:f14} can be written in a general form as
\begin{equation}
	\bm{Ay}+\bm{B}\Delta \bm{x}(\bm{\delta})+\bm{D}\bm{\delta} \leq \bm{J},\ \forall \bm{\delta} \in [-\bm{\delta}^-,\bm{\delta}^+]. \label{j4:robust cstr-g1}
\end{equation}
where $\bm{A,B,D}$ and $\bm{J}$ are constant matrices which can be derived from constraints  \eqref{arwp:f11}--\eqref{arwp:f14}. We note that this general constraint has the same form as presented in \cite{ye2018surrogate}, such that their surrogate affine approximation can be applied. Specifically, we define a constant uncertainty set as
\begin{equation}
	\bm{\Omega}=\{(\bm{\epsilon}^-,\bm{\epsilon}^+)\in \mathbb{R}^{2n_wT}:\bm{0}\leq \bm{\epsilon}^-\leq \bm{1},\bm{0}\leq \bm{\epsilon}^+\leq \bm{1}\},
\end{equation} 
where $\bm{0}$ and $\bm{1}$ are $n_wT \times 1$ column vectors with all zeros and ones, respectively. Additionally, we define $n_w \times n_w$ diagonal matrices $\Theta^-_t=\text{diag}(\bm{\delta}^-_t)$ and $\Theta^+_t=\text{diag}(\bm{\delta}^+_t), t \in \cal T$, with diagonal values equal to elements of the vectors $\bm{\delta}^-_t$ and $\bm{\delta}^+_t$, respectively. Next, we construct the block diagonal matrices
\begin{equation}
\bm{\Theta}^-=\left[\begin{array}{ccc}
	\Theta^-_1  & \bm{0} & \cdots \\
%	\bm{0} & \Theta^-_2 & \bm{0} & \cdots \\
	\vdots  & \ddots & \cdots \\
	\bm{0} & \ddots & \Theta^-_T \\
	\end{array}\right],
\bm{\Theta}^+=\left[\begin{array}{ccc}
\Theta^+_1  & \bm{0} & \cdots \\
%\bm{0} & \Theta^-_2 & \bm{0} & \cdots \\
\vdots  & \ddots & \cdots \\
\bm{0} & \ddots & \Theta^+_T \\
\end{array}\right],	
\end{equation}
where the diagonal elements are square matrices $\Theta^-_t$ and $\Theta^+_t,t \in \cal T$, respectively. Then, according to Lemma 1 of \cite{ye2018surrogate}, it can be shown that the set $\{\bm{\delta}:\bm{\delta} \in [-\bm{\delta}^-,\bm{\delta}^+]\}$ is identical to the image set of $\bm{\Omega}$ under the following operation:
\begin{equation}
	[-\bm{\Theta}^- \ \bm{\Theta}^+]\begin{bmatrix}
	\bm{\epsilon}^{-} \\
	\bm{\epsilon}^{+}
	\end{bmatrix}, \forall  (\bm{\epsilon}^-,\bm{\epsilon}^+)\in  \bm{\Omega}.
\end{equation}
By defining the fixed uncertainty set $\bm{\Omega}$, we consider $\Delta \bm{x}(\bm{\delta})$ to be a surrogate affine function with respect to $(\bm{\epsilon}^-,\bm{\epsilon}^+)$, instead of an affine function of $(\bm{\delta}^-,\bm{\delta}^+)$:
\begin{equation}
	\Delta \bm{x}(\bm{\epsilon})=[\bm{E}^- \ \bm{E}^+]\begin{bmatrix}
	\bm{\epsilon}^{-} \\
	\bm{\epsilon}^{+}
	\end{bmatrix},
\end{equation}
where $\bm{E}^-$ and $\bm{E}^+$ are block diagonal matrix variables to be determined with the following form:
\begin{equation}
\bm{E}^-=\left[\begin{array}{ccc}
\bm{E}^-_1  & \bm{0} & \cdots \\
%	\bm{0} & \Theta^-_2 & \bm{0} & \cdots \\
\vdots  & \ddots & \cdots \\
\bm{0} & \ddots & \bm{E}^-_T \\
\end{array}\right],
\bm{E}^+=\left[\begin{array}{ccc}
\bm{E}^+_1  & \bm{0} & \cdots \\
%\bm{0} & \Theta^-_2 & \bm{0} & \cdots \\
\vdots  & \ddots & \cdots \\
\bm{0} & \ddots & \bm{E}^+_T \\
\end{array}\right].
\end{equation}
Note that the dimension of $\bm{E}^{-}_t$ and $\bm{E}^{+}_t,t \in \cal T$ is $n_f \times n_w$ and the block diagonal structure ensures that the power consumption adjustments of the flexible load aggregators at time $t$ are only related to the wind power deviation at time $t$. Then, we can reformulate constraint \eqref{j4:robust cstr-g1} to be
\begin{align}
\bm{Ay}+\bm{B}[\bm{E}^- \ \bm{E}^+]\begin{bmatrix}
\bm{\epsilon}^{-} \\
\bm{\epsilon}^{+}
\end{bmatrix}+\bm{D}[-\bm{\Theta}^- \ \bm{\Theta}^+]\begin{bmatrix}
\bm{\epsilon}^{-} \\
\bm{\epsilon}^{+}
\end{bmatrix} \leq \bm{J},\nonumber \\
\hspace{4cm} \forall  (\bm{\epsilon}^-,\bm{\epsilon}^+)\in  \bm{\Omega}. \label{j4:robust cstr-g2}
\end{align}
Since the uncertainty set in \eqref{j4:robust cstr-g2} is fixed and from the strong duality, the surrogate affine approximation of \eqref{j4:robust cstr-g1} is given as:
\begin{align}
	& \bm{B}[\bm{E}^- \ \bm{E}^+]+\bm{D}[-\bm{\Theta}^- \ \bm{\Theta}^+]-\bm{\pi} \leq 0, \\
	& \bm{Ay} - \bm{J} +\bm{\pi}\cdot \bm{1} \leq 0,\\
	& \bm{\pi} \geq 0,
\end{align}
which are all linear constraints and $\bm{\pi}$ is the dual variable corresponding to constraints $(\bm{\epsilon}^-,\bm{\epsilon}^+)\in  \bm{\Omega}$. Therefore, the robust constraints \eqref{arwp:f11}--\eqref{arwp:f14} are reformulated into linear form and the original problem can be solved efficiently.

\section{Case Studies}\label{j4:sec4}
	In this section, we apply our model on a modified six-bus transmission network as shown in Fig. \ref{j4:fig1}, with the network information shown in Table \ref{j4:table-2}. The time horizon is 24 hours, so $T=24$. In this network, there are three controllable generators located as buses 1, 2 and 6 and the coefficients of the generation cost function are shown in Table \ref{j4:table-1}. Also, at bus 5, there is a fixed load while one wind generator is located at bus 1. The profiles for the fixed load and the forecast of wind power outputs are shown in Fig. \ref{j4:fig2}. In this figure, we also show the wind power output samples (scaled) with respect to the data of August in 2017 according to CAISO's record \cite{websitecaiso}. These data are used to approximate the integral when calculating the CVaR values. In addition, there are two load aggregators with MDF located at buses 3 and 4. We assume that these two flexible load aggregators have the same flexibility. Specifically, for both of them, we set $x^-_{i,t}=0$ and $x^+_{i,t}=160 \text{ MW}, \forall i, t$. Moreover, the profiles for $\bm{l}_{i}$ and $\bm{u}_{i},i=3,5$, which represent the energy and shifting flexibility, are shown in Fig. \ref{j4:fig3}. As for the CVaR, we set the risk level $\beta=0.9$. Then, we conduct simulations for six cases under different settings with respect to the weighting factors $\eta^1$ and $\eta^2$, as shown in Table \ref{j4:table-3}.

\begin{figure}[!t]
	\centering
	\includegraphics[width=3.2in]{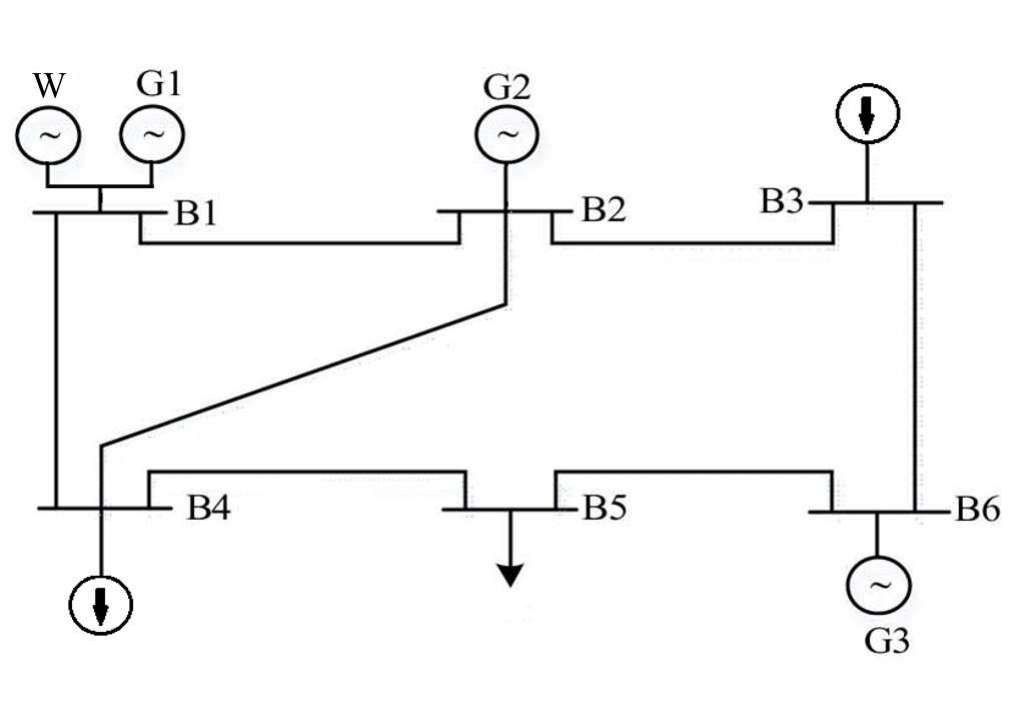}
	\caption{Six-bus transmission network.}
	\label{j4:fig1}
\end{figure}

\begin{table}[!t]
	\caption{Network Information}
	\label{j4:table-2}
	\centering
	\begin{tabular}{cccc}
		\hline
		From Bus & To Bus & Reactance (p.u.)   & Flow Limits (MW)  \\
		\hline
		1 & 2 & 0.170  & 450    \\
		\hline
		1 & 4 & 0.258  & 420    \\
		\hline 
		2 & 3 & 0.037  & 420    \\
		\hline	
		2 & 4 & 0.197  & 450    \\
		\hline 
		3 & 6 & 0.018  & 400    \\
		\hline
		4 & 5 & 0.037  & 400    \\
		\hline		
		5 & 6 & 0.140  & 400    \\
		\hline		
	\end{tabular}
\end{table}
\begin{table}[!t]
	\caption{Generator Data}
	\label{j4:table-1}
	\centering
	\resizebox{\textwidth/2}{!}{%
	\begin{tabular}{cccccc}
		\hline
		Index & $g^-_i(\text{MW})$ & $g^+_i(\text{MW})$   &  $c_i^2(\$/(\text{MW}^2\text{h})$ &  $c_i^1(\$/\text{MWh})$  &  $c_i^0(\$/\text{h})$ \\
		\hline
		G1 & 0 & 1100 &  0.03    & 7      & 0\\
		\hline
		G2 & 0 & 500  & 0.07    & 10     & 0\\
		\hline
		G3 & 0 & 230  & 0.05    & 8      & 0\\
		\hline				
	\end{tabular}
}
\end{table}
\begin{figure}[!t]
	\centering
	\includegraphics[width=3.3in]{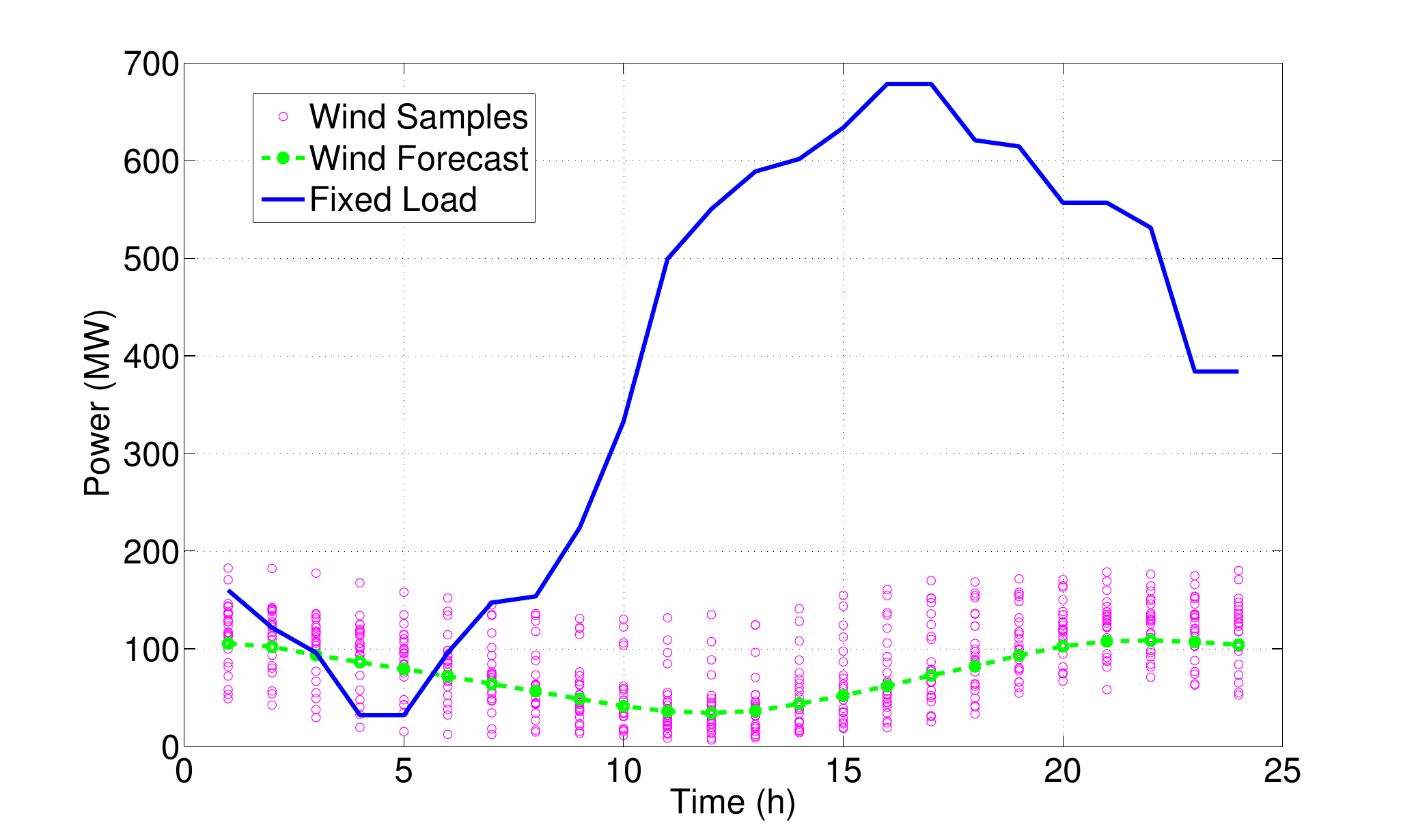}
	\caption{Fixed load, wind power output forecast and wind samples.}
	\label{j4:fig2}
\end{figure}
\begin{figure}[!t]
	\centering
	\includegraphics[width=3.3in]{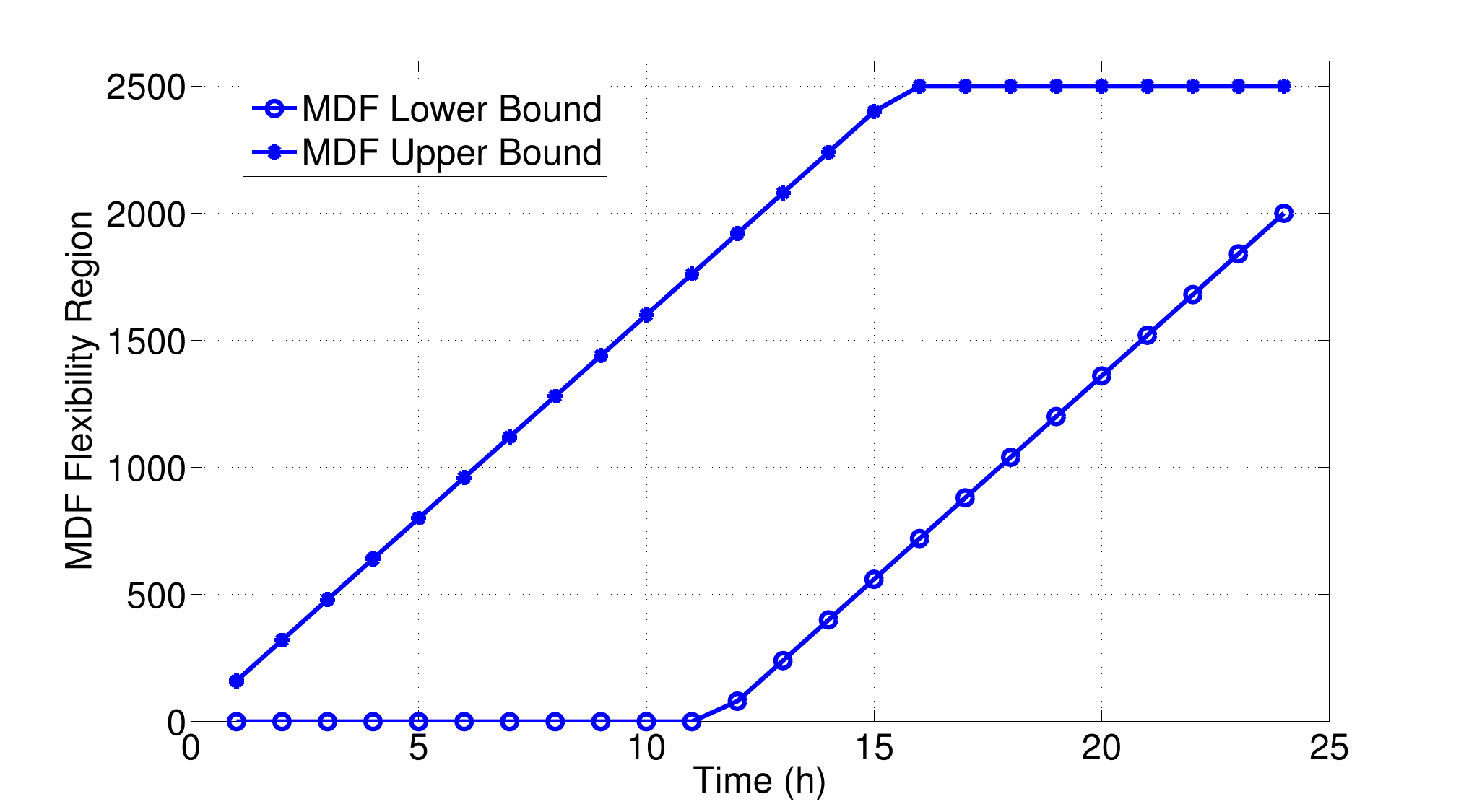}
	\caption{Flexibility region of load aggregators with MDF.}
	\label{j4:fig3}
\end{figure}
\begin{table}[!t]
	\caption{Different Case Settings}
	\label{j4:table-3}
	\centering
	\begin{tabular}{ccccccc}
		\hline
					& Case 1 & Case 2 & Case 3 & Case 4 & Case 5 & Case 6 \\
		\hline
		$\eta^1$    & 10    & 10    & 50    & 100   & 200   & 200  \\
		\hline
		$\eta^2$    & 10    & 100   & 100   & 100   & 100   & 200  \\		\hline				
	\end{tabular}
\end{table}
\begin{figure}[!t]
	\centering
	\includegraphics[width=3.3in]{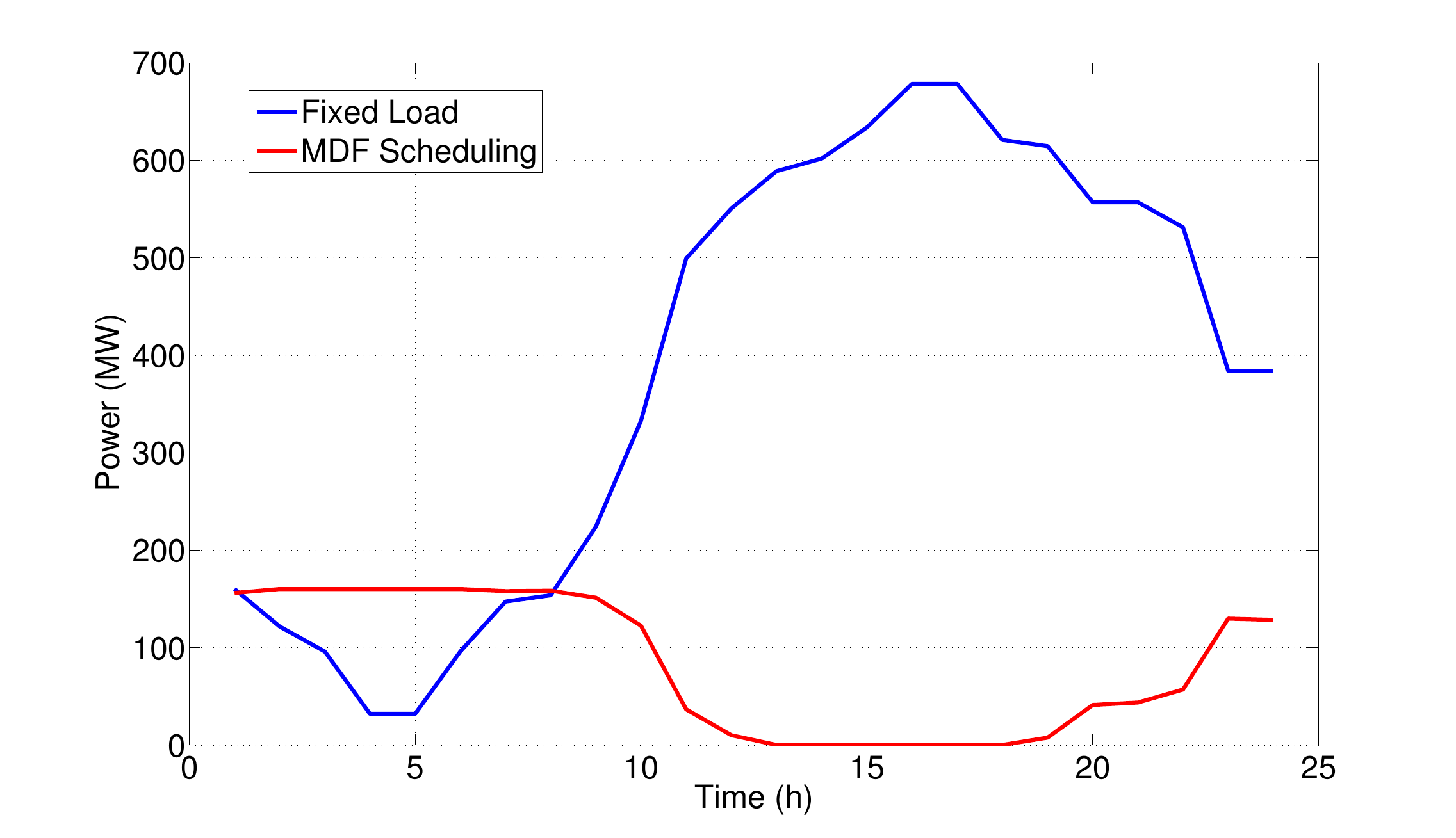}
	\caption{Hourly power consumption of flexible load aggregator at set-points.}
	\label{j4:fig4}
\end{figure}
\begin{figure*}[t!]
	\vspace{-0.0cm}
	\centering     %%% not \center
	\subfigure[Case 1]{\label{j4_fig:3a1}\includegraphics[width=59mm]{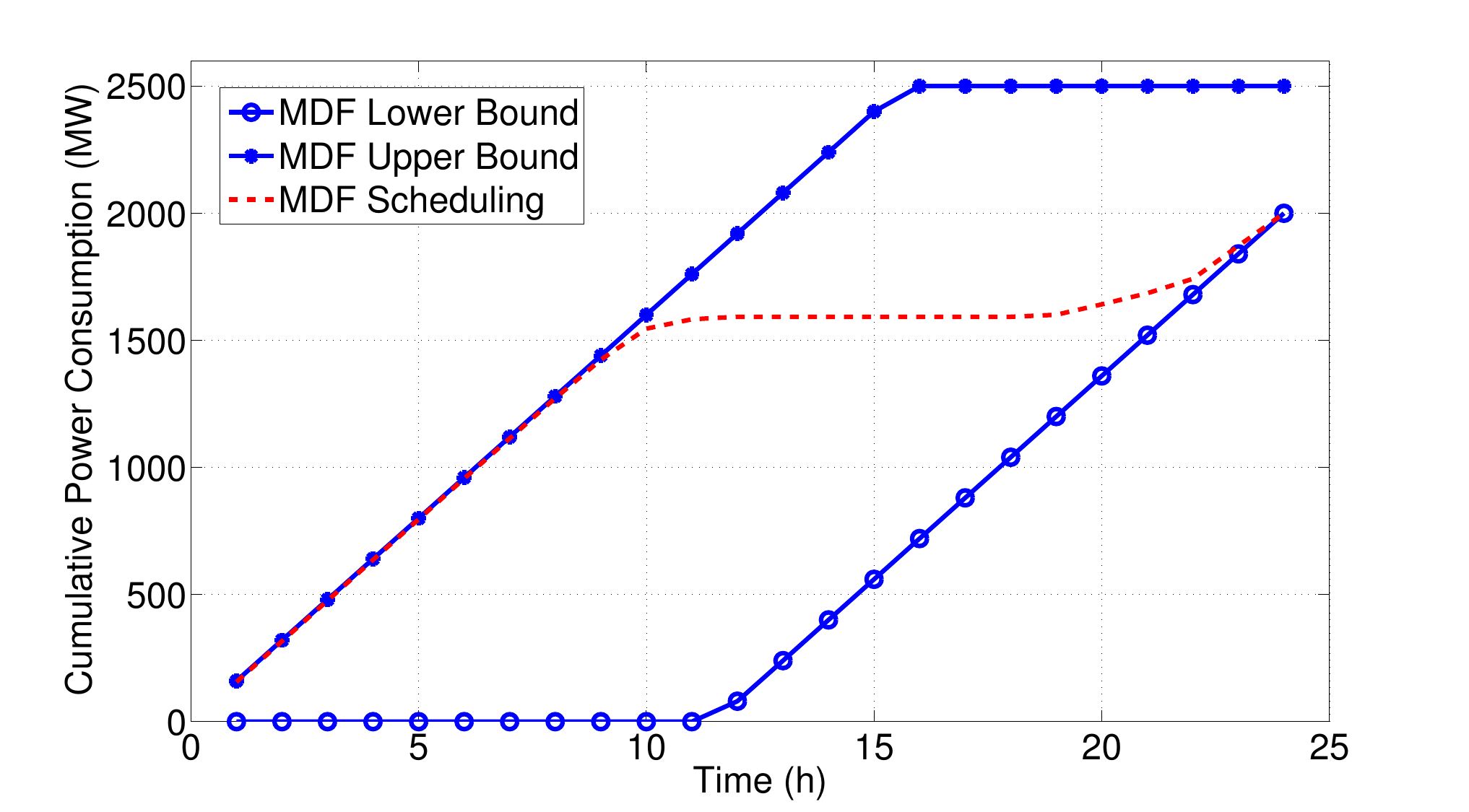}}
	\subfigure[Case 2]{\label{j4_fig:3b1}\includegraphics[width=59mm]{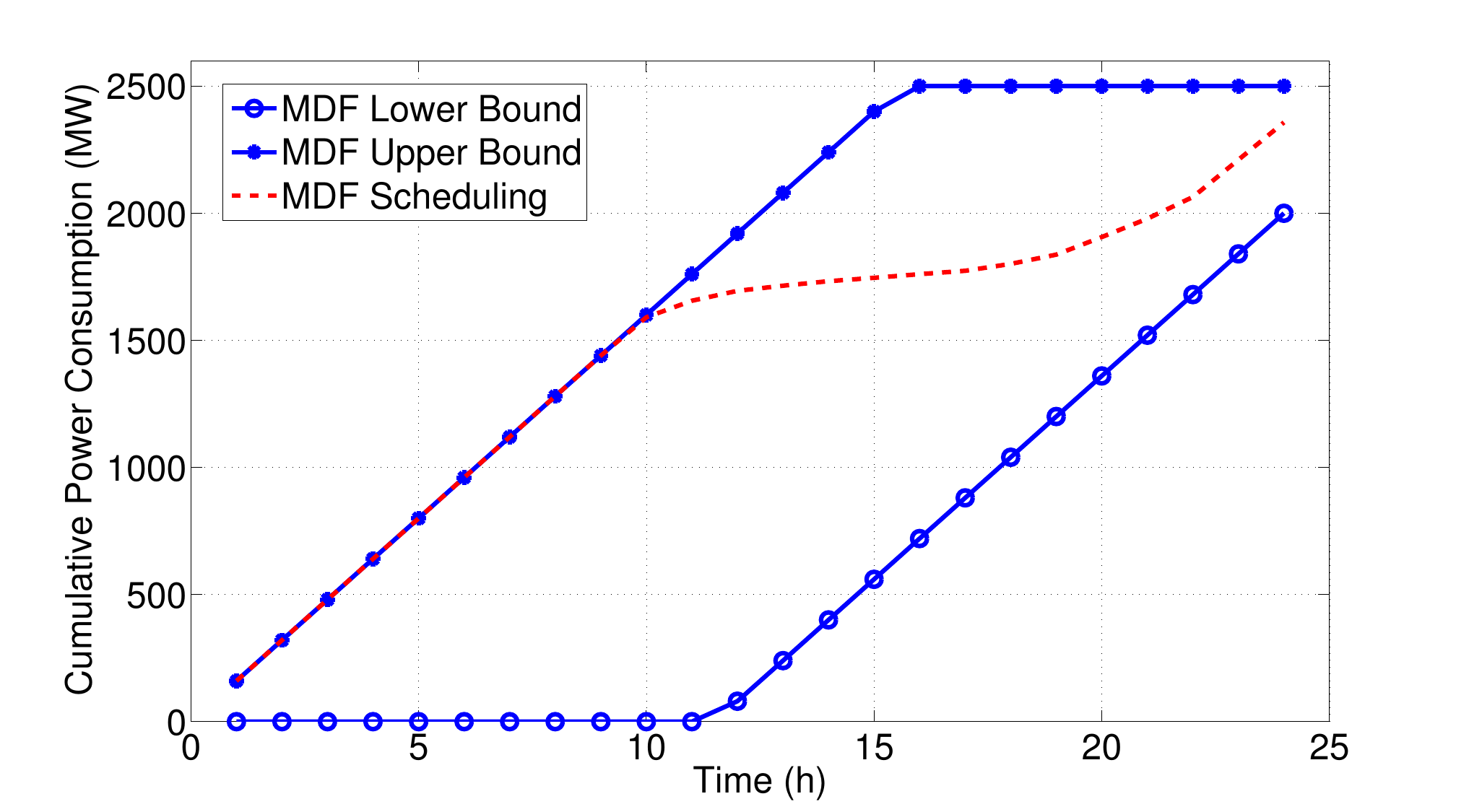}}
	\subfigure[Case 3]{\label{j4_fig:3c1}\includegraphics[width=59mm]{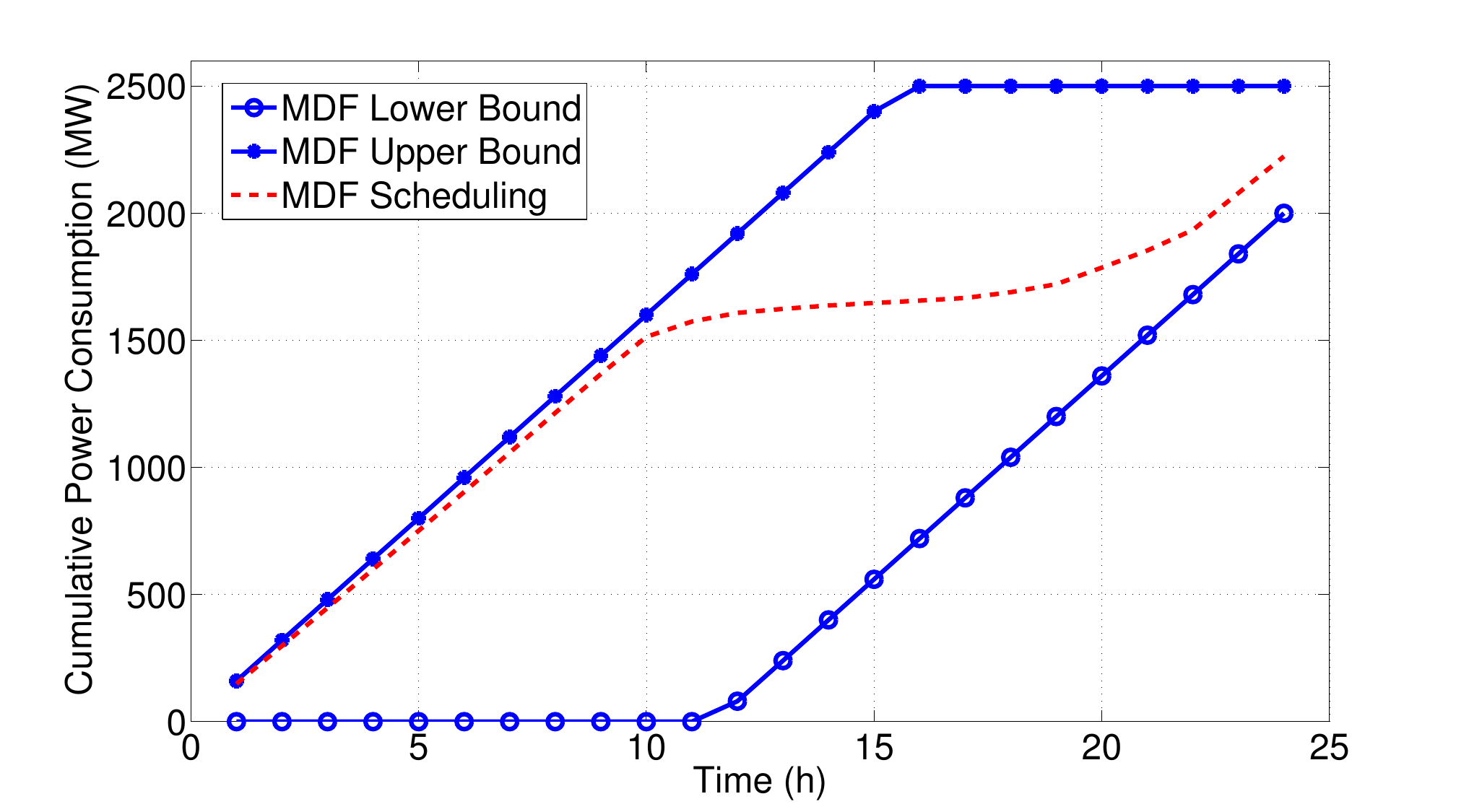}}
	\subfigure[Case 4]{\label{j4_fig:3a2}\includegraphics[width=59mm]{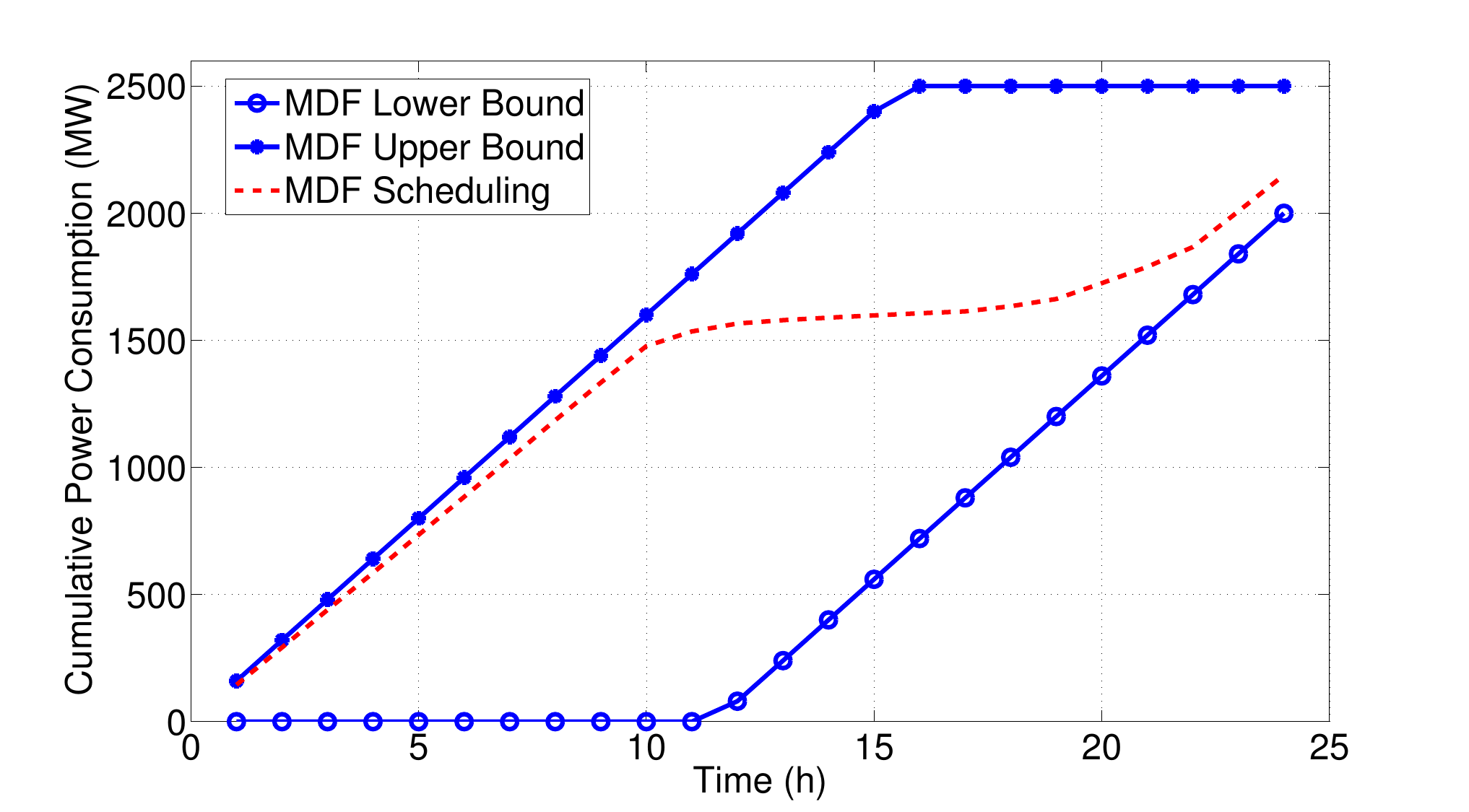}}
	\subfigure[Case 5]{\label{j4_fig:3b2}\includegraphics[width=59mm]{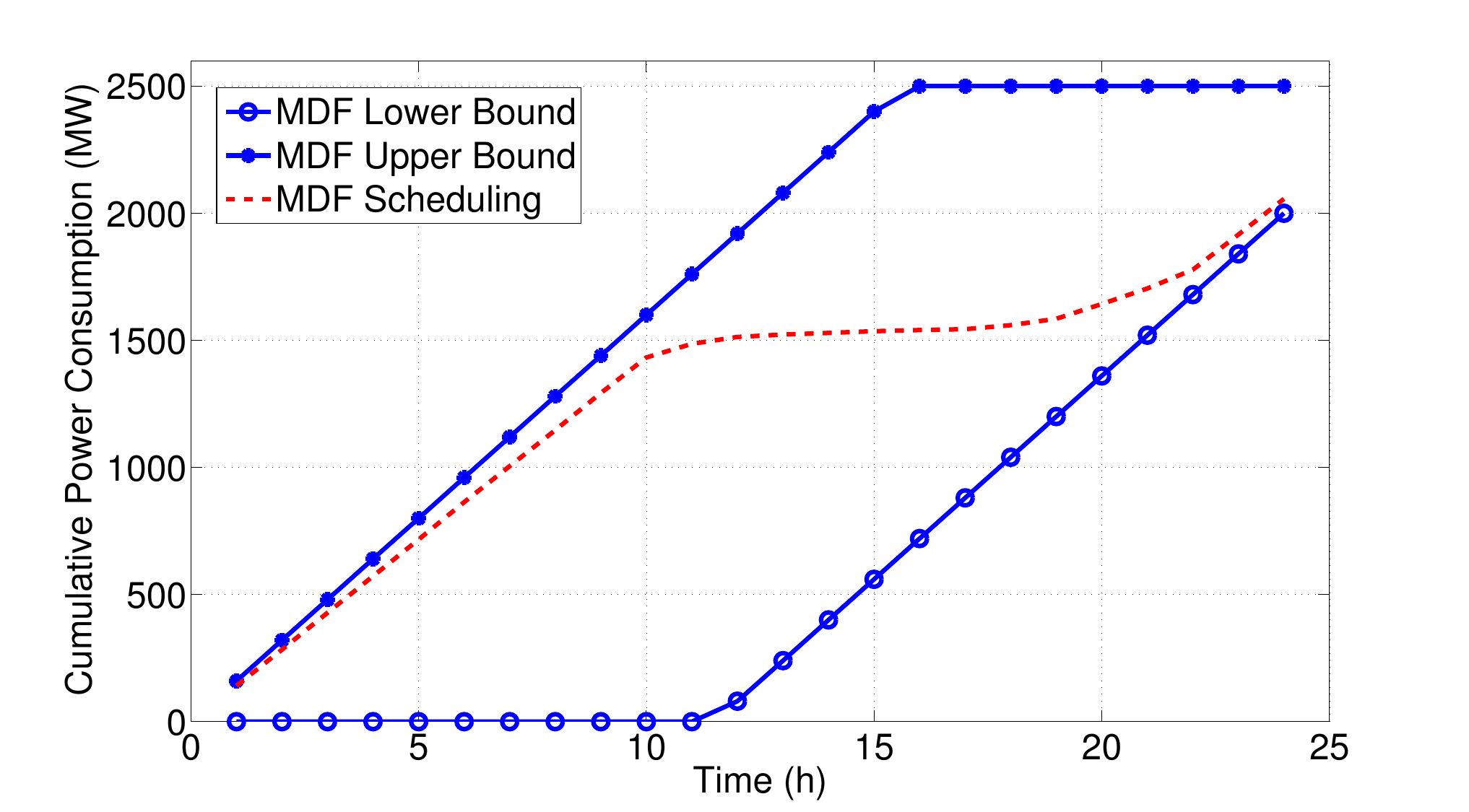}}
	\subfigure[Case 6]{\label{j4_fig:3c2}\includegraphics[width=59mm]{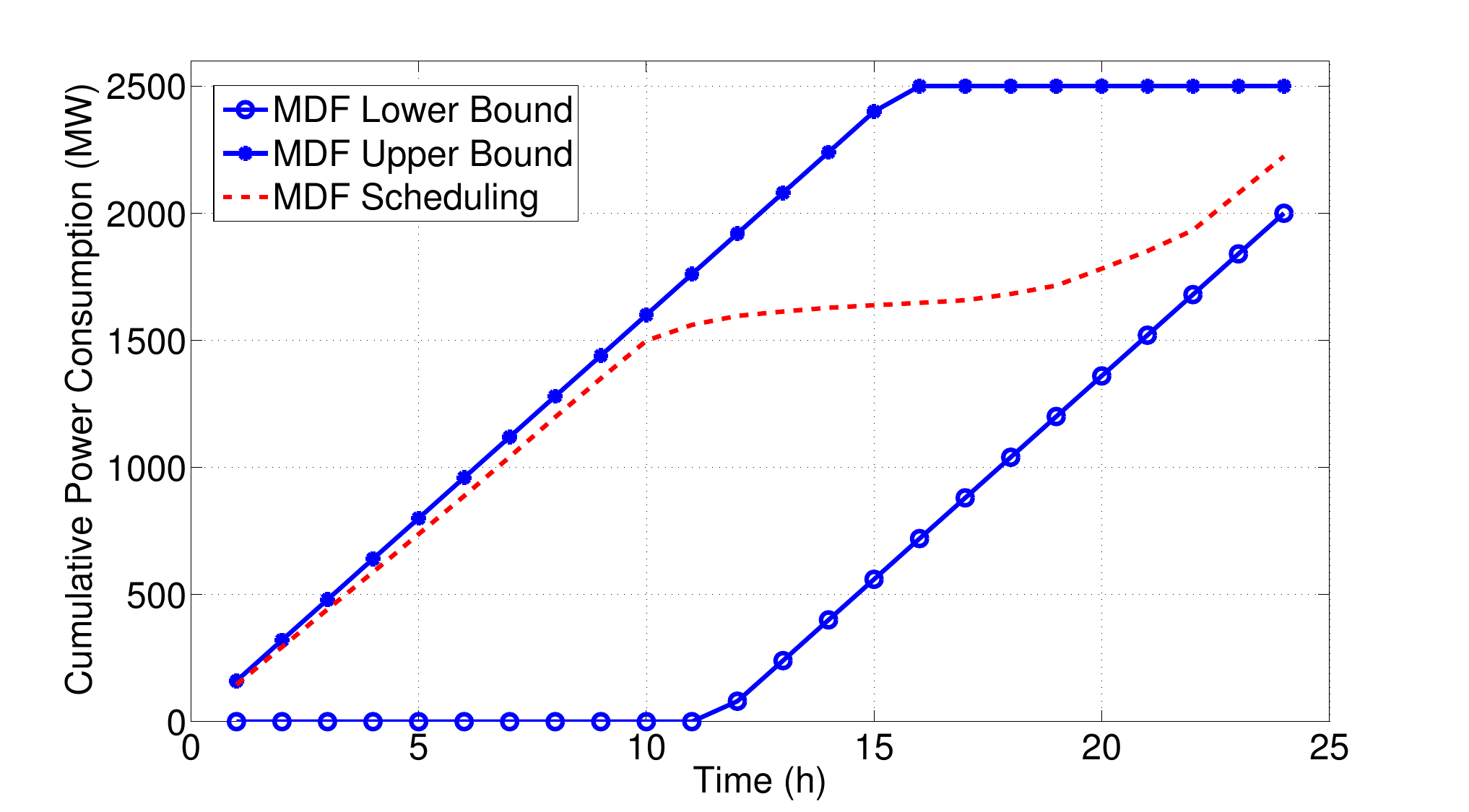}}
	\caption{Cumulative power consumption of flexible load aggregators at set points.}\label{j4_fig:333}
\end{figure*}
\begin{figure*}[t!]
	\vspace{-0.0cm}
	\centering     %%% not \center
	\subfigure[Case 1]{\label{j4_fig:4a1}\includegraphics[width=59mm]{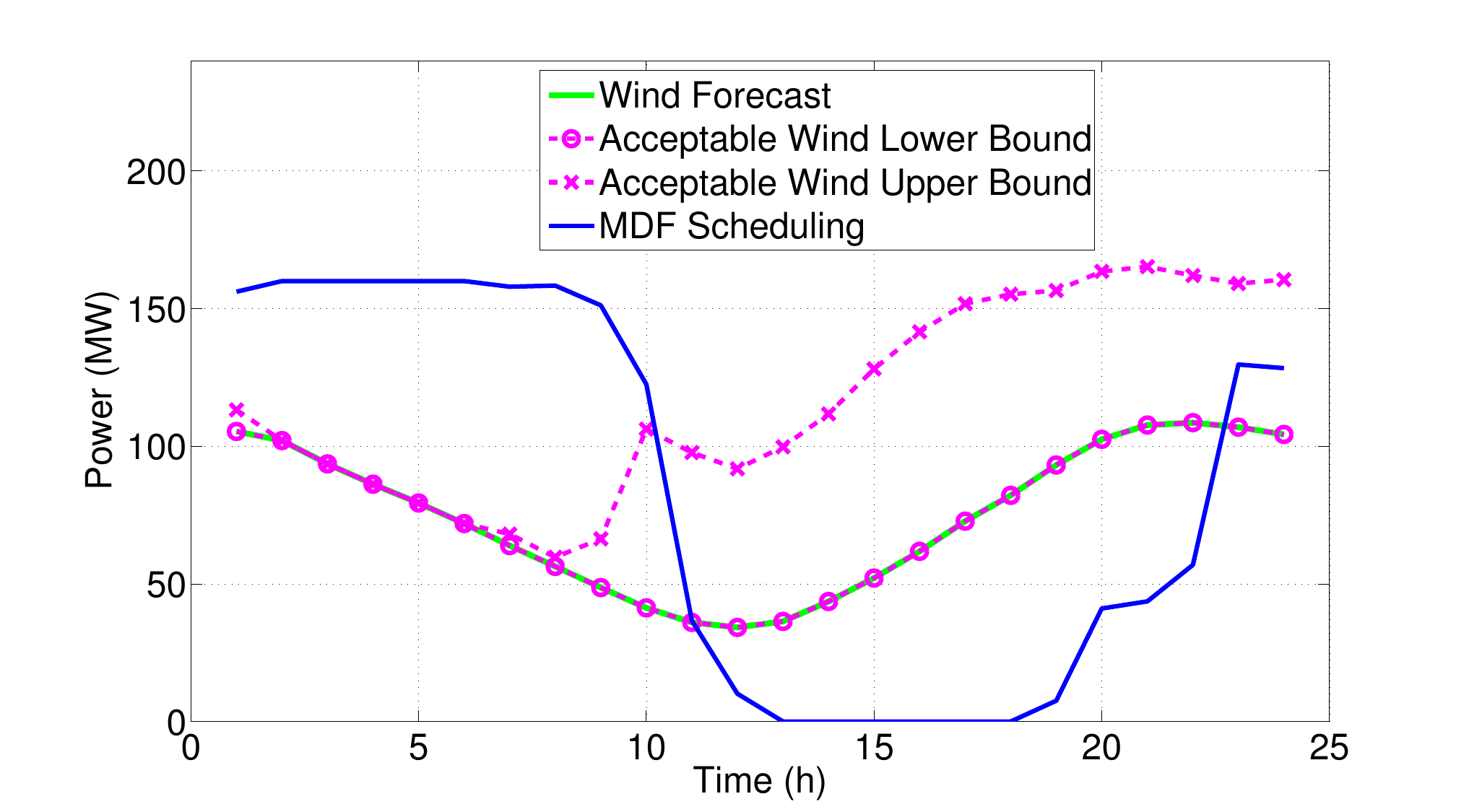}}
	\subfigure[Case 2]{\label{j4_fig:4b1}\includegraphics[width=59mm]{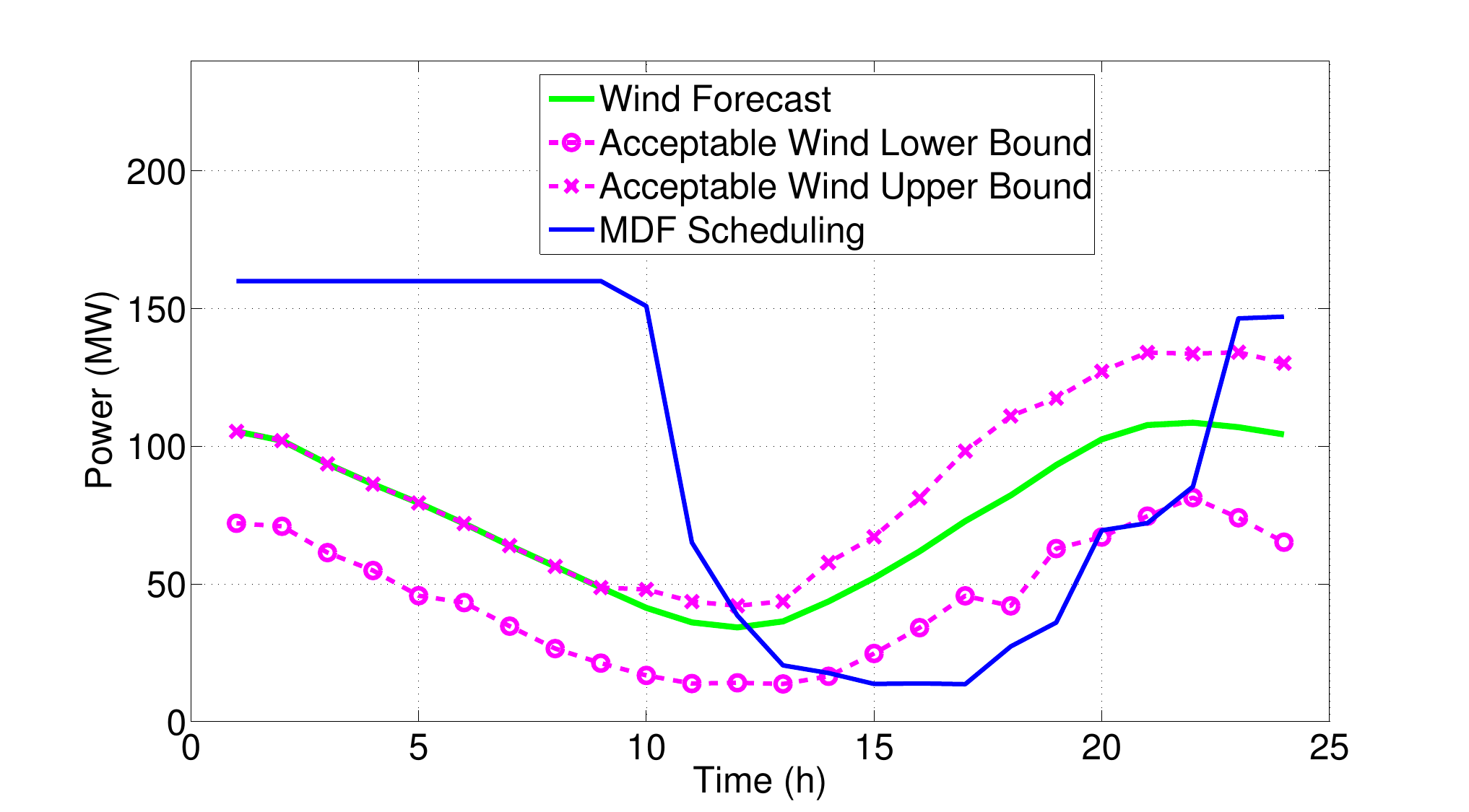}}
	\subfigure[Case 3]{\label{j4_fig:4c1}\includegraphics[width=59mm]{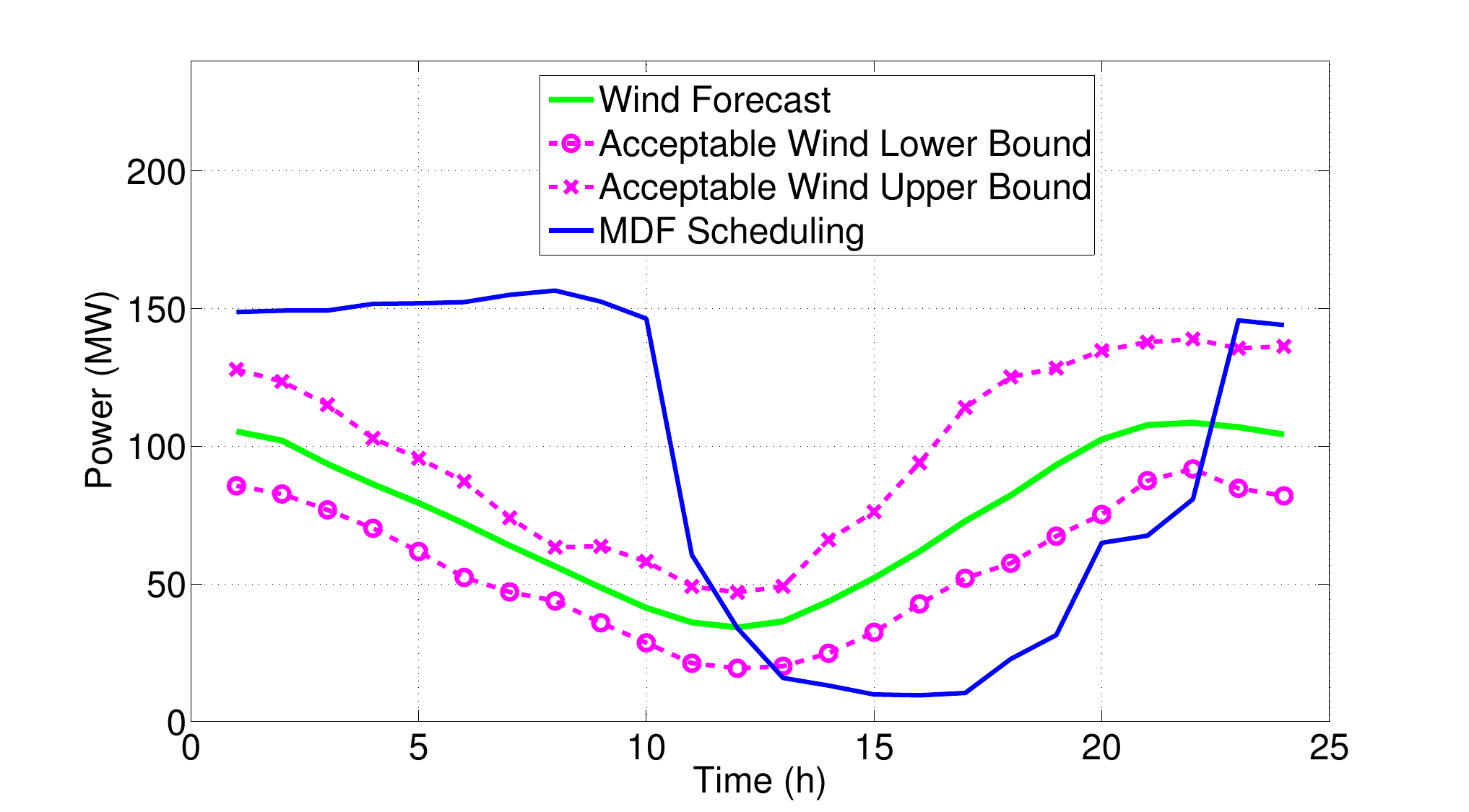}}
	\subfigure[Case 4]{\label{j4_fig:4a2}\includegraphics[width=59mm]{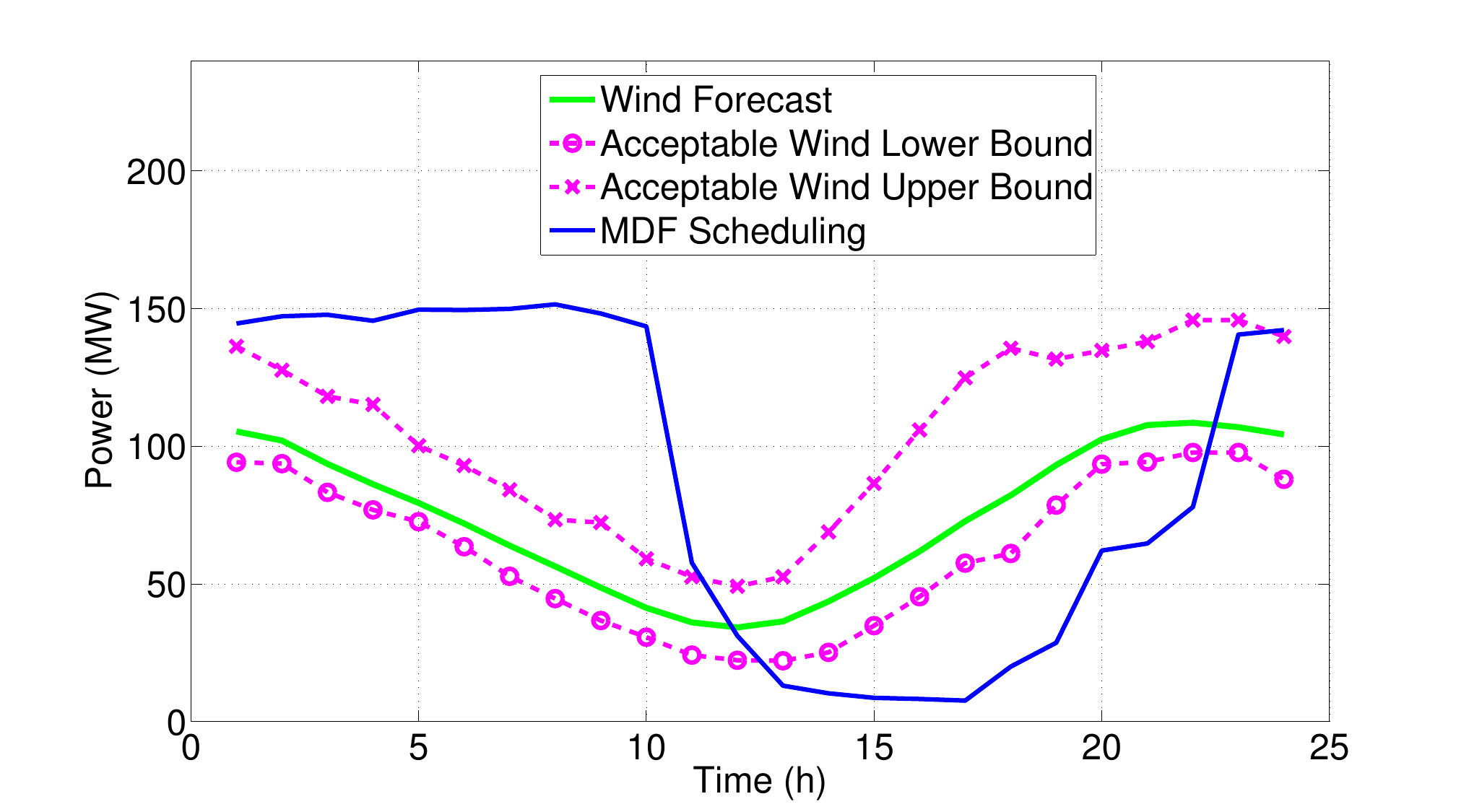}}
	\subfigure[Case 5]{\label{j4_fig:4b2}\includegraphics[width=59mm]{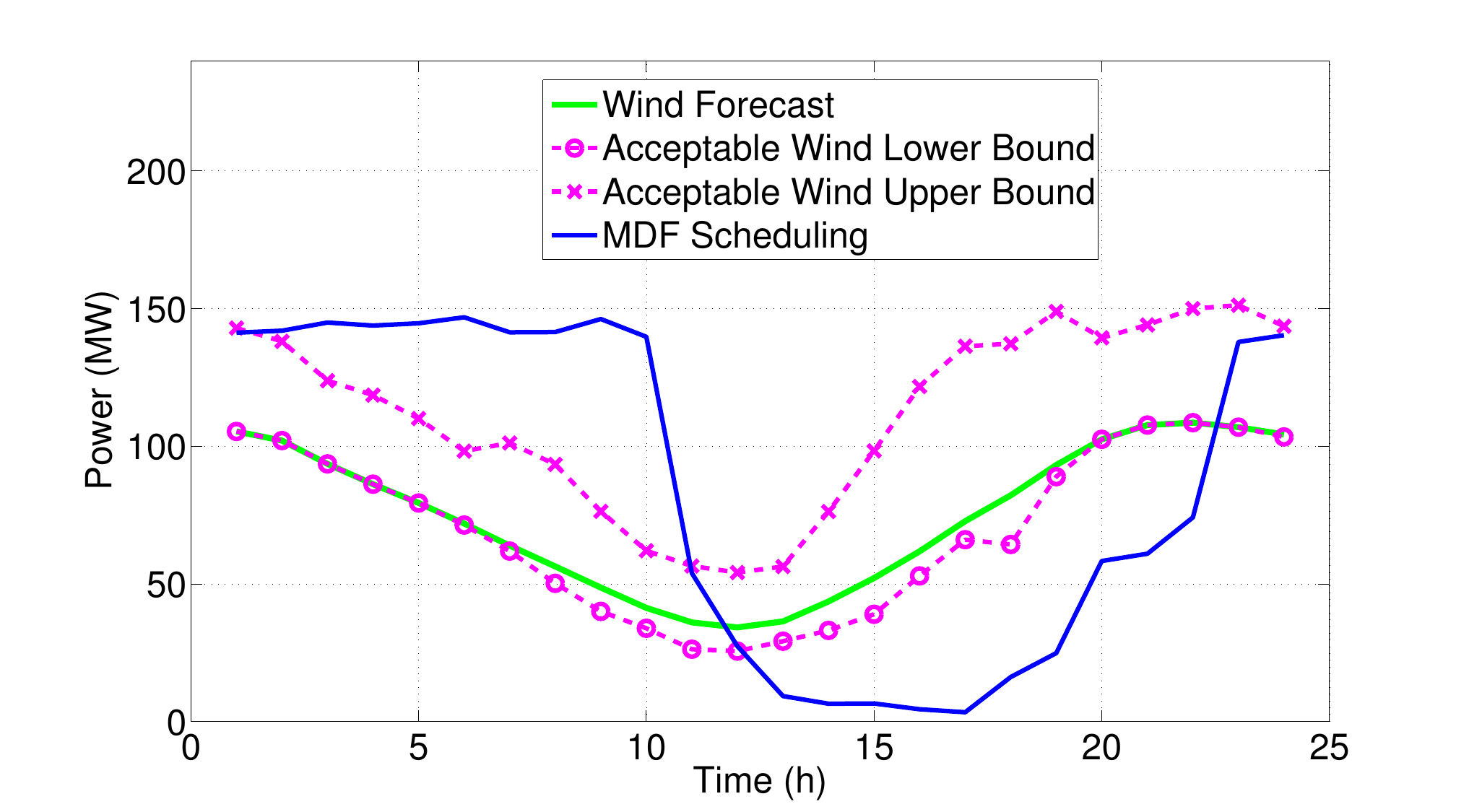}}
	\subfigure[Case 6]{\label{j4_fig:4c2}\includegraphics[width=59mm]{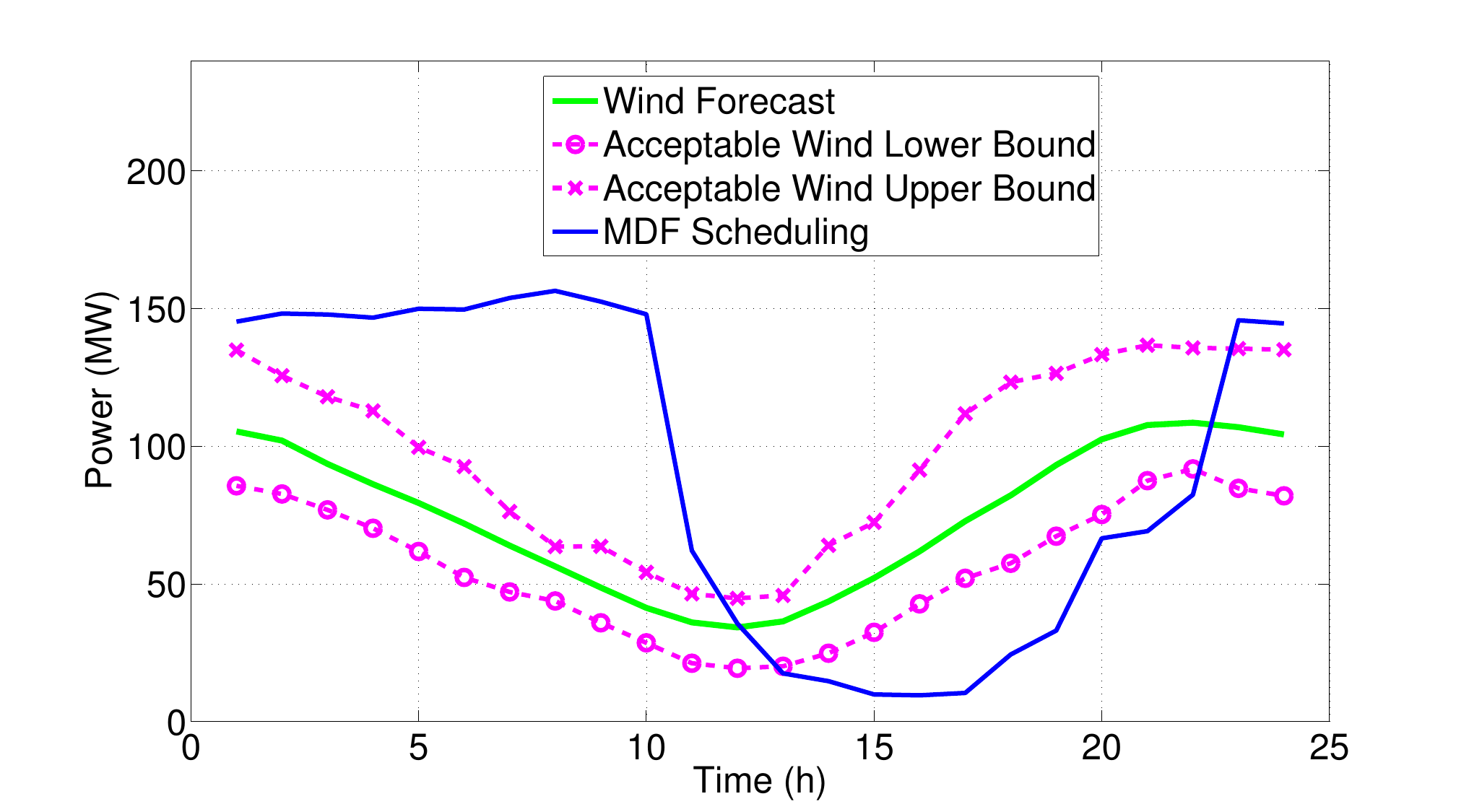}}
	\caption{Admissible region of wind power uncertainty set.}\label{j4_fig:444}
\end{figure*}

First, we show the hourly power consumption scheduling at the set points of the flexible load aggregators (i.e., the profile of $\overline{\bm{x}}_{\cdot i},i=3,4$) in Fig. \ref{j4:fig4} under the setting of Case 1. Since we find that there is no congestion after the problem is solved, the scheduling of the two load aggregators are identical and we just need to show the result for one of them. From Fig. \ref{j4:fig4}, we observe that the scheduling of the flexible load has a complementary profile compared with the fixed load profile during the peak load period (from 8:00 to 23:00). This is reasonable since such scheduling will help to flatten the overall load profile, which can result in a lower generation cost. We also note that during the off-peak period, the flexible load scheduling is not complementary to the fixed load. In fact, it reaches the maximum possible power rate (160 MW). This is because although the load aggregator has some flexibility, it still has a minimum energy consumption requirement to meet ($l_{i,t},t=24$). Therefore, the red curve in Fig. \ref{j4:fig4} shows how the flexible load aggregator helps reduce the generation cost while meeting its own constraints. Note that the similar results are observed for Case 2 to Case 6, which will be shown in the following.

\begin{table}[!t]
	\caption{Simulation Results For CVaR and Generation Cost}
	\label{j4:table-4}
	\centering
	\begin{tabular}{ccccccc}
		\hline
		& Case1 & Case2 & Case3 & Case4 & Case5 & Case6 \\
		\hline		  	
		$\phi^1$ 	& 145.0    & 272.3    & 170.2    & 133.5   & 97.6   & 166.9  \\
		\hline
		$\phi^2$ 	& 394.5 & 91.2 & 177.8 & 238.4 & 326.4 & 177.8 \\
		\hline
		$C^G(\times 10^5)$        & 1.757 & 1.933 & 1.875 & 1.843 & 1.802 & 1.878 \\
		\hline	
		$C^G_{w/o}(\times 10^5)$  & 1.757 & 2.099 & 1.979 & 1.913 & 1.829 & 1.982 \\
		\hline						
	\end{tabular}
\end{table}

Next, we show the cumulative energy consumption of flexible load aggregators at set points (i.e., $\bm{L}\overline{\bm{x}}_{\cdot i}$) in Fig. \ref{j4_fig:333} and the admissible lower and upper bounds of the wind power uncertainty set are shown in Fig. \ref{j4_fig:444}. Moreover, we show the total CVaR values ($\phi^1=\sum_{i\in \cal{N}^W}\sum_{t\in \cal T}\phi^{1,\beta}_{i,t}$, $\phi^2=\sum_{i\in \cal{N}^W}\sum_{t\in \cal T}\phi^{2,\beta}_{i,t}$) and generation cost ($C^G=\sum_{i\in\cal{N}^G}\sum_{t\in \cal T}C^G(g_{i,t})$) under different settings in Table \ref{j4:table-4}. 

First, we focus on Figs. \ref{j4_fig:3a1} and \ref{j4_fig:4a1} for Case 1. We can see that since the weighting factors are small, the flexible load consumes power at the maximum rate during the off peak period and consumes little during the peak load period. Also, it reaches the lowest possible point at $t=24$. In this way, the generation cost can be reduced significantly and the CVaR is not important in this case. However, the upper bound for the wind power uncertainty set is high, because this can be achieved with no influence on the low generation cost viewing that there is a large space between the power scheduling of the flexible load and its upper bound in Fig. \ref{j4_fig:3a1}. By contrast, there is no space for the downward deviation of the wind power output since the lower bound is already reached.

When we gradually increase the importance of CVaR with larger weighting factors (Case 2 to Case 6), we observe that the admissible region for the wind power uncertainty set becomes larger. This is because a larger uncertainty set can result in a lower CVaR value. From Fig. \ref{j4_fig:333}, the flexible load scheduling curve does not always reach the lowest point at $t=24$. This shows that the low generation cost is sacrificed for a low CVaR value. Also, there exists certain space between the red curve and both the upper and lower bounds. Consequently, both the upward and downward deviation of the wind power outputs can be handled by the load aggregators with MDF as shown in Fig. \ref{j4_fig:4b1}--Fig. \ref{j4_fig:4c2}. In addition, we note that although $\eta^1$ and $\eta^2$ are non-decreasing from Case 1 to Case 6, it does not always result in a non-decreasing variation for the lower and upper bounds of the wind power uncertainty set. This is because the relative ratio between $\eta^1$ and $\eta^2$ is also of great importance. As the flexibility region of the load aggregator is bounded from both above and below, the increase of the admissible region for the wind power deviation in one direction (upward or downward) will lead to the decrease of that in the other direction. This can also be seen from Table \ref{j4:table-4}. We note that the total CVaR values for the wind curtailment ($\phi^1$) and power deficiency ($\phi^2$) are not always decreasing. Typically, the increase of one usually comes with the decrease of the other. In addition, another case is considered when no flexible load is available. In this case, the flexible load consumption is fixed to the red curve as shown in Fig. \ref{j4_fig:333}. Also, the wind uncertainty region is the area between the two purple curves as shown in Fig. \ref{j4_fig:444}, and it will be handled by the controllable generators. Then, the worst case generation costs under wind uncertainty without flexible load are shown in the last row of Table \ref{j4:table-4}. We observe that if no flexible load is involved, the generation cost will be higher and this shows the benefit of load aggregators with MDF for reducing the generation cost.

Therefore, we have demonstrated that the MDF from the flexible load aggregators can be leveraged by the system operator to handle the uncertainty of wind power output and lower down the generation cost. More importantly, the uncertainty set that can be handled by the system is also co-optimized considering the balance between the generation cost and the CVaR related system risks while satisfying all the operational constraints of generators and load aggregators.

%\begin{figure}[!t]
%	\centering
%	\includegraphics[width=2.5in]{fig/pic-6-bus-1.jpg}
%	\caption{Six-bus transmission network.}
%	\label{mdf_fig:121}
%\end{figure}
%\begin{table}[!t]
%	\caption{Generator Data}
%	\label{mdf_table:121}
%	\centering
%	\begin{tabular}{ccccccc}
%		\hline
%		Index & $P_G^{min} $(MW) & $P_G^{max} $(MW) &   a(\$/$\textrm{MW}^2\textrm{h}$) & b(\$/MWh) & c(\$) \\
%		\hline
%		G1 & 40 & 220 &  0.03 & 7 & 100   \\
%		\hline
%		G2 & 10 & 200 &  0.07 & 10 & 104 \\
%		\hline
%		G3 & 0  & 25 & 0.05 & 8 & 110    \\
%		\hline
%	\end{tabular}
%\end{table}

%\begin{figure*}[t!]
%		\centering     %%% not \center
%		\subfigure[Cleared $|\alpha^{E1}|$ at Bus 5.]{\label{mdf_fig:3a1}\includegraphics[width=58mm]{fig/alpE1_1_v2.eps}}
%		\subfigure[Cleared $|\alpha^{E1}|$ at Bus 7.]{\label{mdf_fig:3b1}\includegraphics[width=58mm]{fig/alpE1_2_v2.eps}}
%		\subfigure[Cleared $|\alpha^{E1}|$ at Bus 9.]{\label{mdf_fig:3c1}\includegraphics[width=58mm]{fig/alpE1_3_v2.eps}}
%		\subfigure[Cleared $|\alpha^{P2}|$ at Bus 5.]{\label{mdf_fig:3a2}\includegraphics[width=58mm]{fig/alpP2_1_v2.eps}}
%		\subfigure[Cleared $|\alpha^{P2}|$ at Bus 7.]{\label{mdf_fig:3b2}\includegraphics[width=58mm]{fig/alpP2_2_v2.eps}}
%		\subfigure[Cleared $|\alpha^{P2}|$ at Bus 9.]{\label{mdf_fig:3c2}\includegraphics[width=58mm]{fig/alpP2_3_v2.eps}}
%		\caption{Market clearing results for $\gamma^P$ ranging from 110 to 910 and $\gamma^E$ ranging from 100 to 2500.}\label{mdf_fig:333}
%\end{figure*}

\section{Conclusions}\label{j4:sec5}
In this paper, we propose a model for the co-optimization of the wind power uncertainty set and the scheduling of load aggregators with multi-dimensional demand flexibility, such that the system operator can find a balance between the low generation cost and system risks. The system loss associated with the wind curtailment and the power deficiency are captured by the CVaR values. Also, we have successfully transformed the calculation of CVaR values into solving a linear program and the robust constraints with decision variables in the uncertainty set is handled by the surrogate affine approximation method. We applied our model on a six-bus transmission network and demonstrate that the multi-dimensional demand flexibility from the flexible load aggregators can help the system to handle the wind power uncertainty while reducing the generation cost.

\section*{Acknowledgement}
This work was supported by the Hong Kong Research Grants Councils General Research Fund under Project 16210215.

% Note that the IEEE does not put floats in the very first column
% - or typically anywhere on the first page for that matter. Also,
% in-text middle ("here") positioning is typically not used, but it
% is allowed and encouraged for Computer Society conferences (but
% not Computer Society journals). Most IEEE journals/conferences use
% top floats exclusively. 
% Note that, LaTeX2e, unlike IEEE journals/conferences, places
% footnotes above bottom floats. This can be corrected via the
% \fnbelowfloat command of the stfloats package.

%
%\section{Conclusion}
%The conclusion goes here.
%

% or


%\appendices
%% you can choose not to have a title for an appendix
%% if you want by leaving the argument blank
%\section{}

%\section*{Acknowledgment}

\ifCLASSOPTIONcaptionsoff
  \newpage
\fi

\end{document}